\documentclass[prd, twocolumn,superscriptaddress, notitlepage]{revtex4-1}
\usepackage{xcolor}
\usepackage{graphicx}
\usepackage{wrapfig}
\usepackage{amsmath}
\usepackage{amsfonts}
\usepackage{amssymb}
\usepackage{multirow}
\usepackage{slashed}
\usepackage{physics}
\usepackage{float}
\usepackage{dcolumn}
\usepackage[colorlinks=true, linkcolor=blue, citecolor=blue, urlcolor=blue]{hyperref}

\newcommand\befs{\begin{figure*}}
\newcommand\eefs[1]{\label{fig:#1}\end{figure*}}
\newcommand\bef{\begin{figure}}
\newcommand\eef[1]{\vskip -0.125cm \label{fig:#1}\end{figure}}
\newcommand\beq{\begin{equation}}
\newcommand\eeq[1]{\label{#1}\end{equation}}
\newcommand\beqa{\begin{eqnarray}}
\newcommand\eeqa[1]{\label{#1}\end{eqnarray}}
\newcommand\bet{\begin{table}}
\newcommand\eet[1]{\label{tb:#1}\end{table}}
\newcommand\bets{\begin{table*}}
\newcommand\eets[1]{\label{tb:#1}\end{table*}}
\newcommand{\be}{\begin{equation}}
\newcommand{\ee}{\end{equation}}
\newcommand{\bea}{\begin{eqnarray}}
\newcommand{\eea}{\end{eqnarray}}
\newcommand\fgn[1]{Fig.\ \ref{fig:#1}}
\newcommand\eqn[1]{Eq.\ (\ref{#1})}

\newcommand\tbn[1]{Table \ref{tb:#1}}

\begin{document}

\widetext

\title{QED$_3$-inspired three-dimensional conformal lattice gauge theory without fine-tuning}

\author{Nikhil\ \surname{Karthik}}
\email{nkarthik.work@gmail.com}
\affiliation{Physics Department, Brookhaven National Laboratory, Upton, New York 11973-5000, USA}
\author{Rajamani\ \surname{Narayanan}}
\email{rajamani.narayanan@fiu.edu}
\affiliation{Department of Physics, Florida International University, Miami, FL 33199}

\begin{abstract}

We construct a conformal lattice theory with only gauge degrees of
freedom based on the induced non-local gauge action in QED$_3$
coupled to large number of flavors $N$ of massless two-component
Dirac fermions. This lattice system displays signatures of criticality
in gauge observables, without any fine-tuning of couplings and can
be studied without Monte Carlo critical slow-down.  By coupling
exactly massless fermion sources to the lattice gauge model, we
demonstrate that non-trivial anomalous dimensions are induced in
fermion bilinears depending on the dimensionless electric charge
of the fermion.  We present a proof-of-principle lattice computation
of the Wilson-coefficients of various fermion bilinear three-point
functions.  Finally, by mapping the charge $q$ of fermion in the
model to a flavor $N$ in massless QED$_3$, we point to an universality
in low-lying Dirac spectrum and an evidence of self-duality of $N=2$
QED$_3$.

\end{abstract}

\date{\today}
\maketitle

{\sl Introduction. --}
Extraction of conformal field theory (CFT) data plays an important
role in our understanding of critical phenomena.  An important set
of conformal data are the scaling dimensions of operators that
classify the relevant and irrelevant operators in a CFT. This data
can be used to abstract the source of dynamical scale-breaking in
the long-distance limit of quantum field theories in terms of few
symmetry-breaking operators that turn relevant.  The operator product
expansion (OPE) coefficients in the CFT correlation functions are
another set of highly constrained conformal data.  The formal
structure of CFT and its data has been explored over decades and
one can refer~\cite{DiFrancesco:1997nk} for a survey of the
subject;~\cite{Osborn:1993cr} for a discussion not restricted to
two dimensions
and~\cite{Rychkov:2016iqz,Simmons-Duffin:2016gjk,Poland:2018epd}
for recent developments in dimensions greater than two.  Monte Carlo
(MC) studies of strongly interacting CFTs are difficult owing to a
combined effect of the required precise tuning of couplings, an
increase in MC auto-correlation time closer to a critical point and
the need for large system sizes. Notwithstanding such difficulties,
the CFT data in many bosonic spin systems have been extracted from
traditional MC (e.g.,~\cite{Banerjee:2017fcx,Banerjee:2019jpw} for
recent determinations in 3d O($N$) models) as well as using radial
lattice quantization~\cite{Brower:2012vg,Neuberger:2014pya,Brower:2020jqj}.
At present, however, three-dimensional fermionic CFTs have been of
great interest, particularly owing to recent works related to
dualities~\cite{Son:2015xqa,Seiberg:2016gmd,Karch:2016sxi}, and
therefore, MC based search for three-dimensional fermionic CFTs
(such
as,~\cite{Hands:2004bh,Karthik:2018nzf,Xu:2018wyg,Hands:2018vrd,Wellegehausen:2017goy,Chandrasekharan:2011mn})
is of paramount importance.

One such three-dimensional interacting fermionic CFT is approached
in the infrared limit of the parity-invariant noncompact quantum
electrodynamics (QED${}_3$) with $N$ (even) flavors of massless two-component Dirac
fermions in the limit of large-$N$; to leading order, the effect
of fermion is to convert the $p^{-2}$ Maxwell photon propagator
into a conformal $16(N g^2 p)^{-1}$ photon
propagator~\cite{Appelquist:1986qw} in the limit of small momentum
$p$, where $g^2$ is the dimensionful Maxwell coupling.  This suggests
replacing the usual Maxwell action for the gauge field $A_\mu$ by
a conformal gauge action~\cite{Giombi:2015haa}
\be
S_g = \frac{1}{q^2}\int \frac{d^3 p}{(2\pi)^3} A_\mu(p)  \left(\frac{p^2 \delta_{\mu\nu} - p_\mu p_\nu}{p}\right) A_\nu(-p),\label{cqed}
\ee
with a dimensionless coupling $q^2(N)=32/N$ for large-$N$, thereby
obtaining results consistent with an interacting conformal field
theory in a $\frac{1}{N}$ expansion. The conformal nature of the
above quadratic action can be seen in the tensorial structure of
$n$-point functions of field strength $F_{\mu\nu}$ that is consistent
with conformal symmetry~\cite{Giombi:2015haa,ElShowk:2011gz}.  Since
the dimension of $F_{\mu\nu}$ is fixed by gauge-invariance, it is
only for the $1/p$ kernel of the above quadratic action, the coupling
becomes dimensionless in three-dimensions.
Both~\cite{Appelquist:1986qw,Giombi:2015haa} approaches  are
consistent with a scale invariant field theory only if $N$ is above
some critical value but recent numerical
analyses~\cite{Karthik:2015sgq,Karthik:2016ppr} of QED${}_3$ have
shown that the theory likely remains scale- (or conformal-) invariant
all the way down to the minimum $N=2$.  This suggests that the
induced gauge action from the fermion is conformal for any non-zero
$N$, and it might be possible to capture  many aspects of the
infrared physics of QED$_3$ by appropriately modeling this induced
non-local action --- we do so by using the quadratic conformal gauge
action \eqn{cqed}, however with an otherwise unknown $q$-$N$ relation,
$q^2(N)$, which for general $N$ needs to be determined from first
principles, and assuming that effect of terms in the induced-action
which are higher-order in gauge field are negligible. This motivated
us to consider the action in \eqn{cqed} in its own right as an
interacting CFT for any $q^2$ obtained without
tuning any couplings, and probed by massless spectator fermions.
It is the primary aim of this letter to use a lattice regularization
of \eqn{cqed} and show that this CFT induces non-trivial conformal
data in fermionic observables depending on the value of $q$, thereby
making it a powerful model system for lattice studies of fermion
CFTs.  Finally, we will close the loop and demonstrate numerically
that this conformal gauge theory for arbitrary $q^2$ probed by
spectator fermions can describe universal features in a corresponding
$N$-flavor QED$_3$.

{\sl The model and signatures of its criticality in pure-gauge observables --}
The noncompact U(1) lattice gauge model we consider is the regularized
version of \eqn{cqed} on $L^3$ periodic lattice, given by
\beqa
&&Z=\left(\prod_{x,\mu}\int_{-\infty}^\infty d\theta_\mu(x)\right) e^{-S_g(\theta)},\ \text{with}\cr
&& S_g = \frac{1}{2} \sum_{\mu,\nu=1}^3\sum_{x,y} F_{\mu\nu}(x)\Box^{-1/2}(x,y) F_{\mu\nu}(y),
\eeqa{partition}
where $\theta_\mu(x)$ are real-valued gauge fields that reside on
the links connecting site $x$ to $x+\hat \mu$, with a field strength
$F_{\mu\nu}=\Delta_\mu\theta_\nu(x)-\Delta_\nu\theta_\mu(x)$ where
$\Delta_\mu$ is the discrete forward derivative. The three-dimensional
discrete Laplacian is $\Box=\sum_{\mu} \Delta_\mu \Delta^\dagger_\mu$.
The model lacks any tunable dimensionful parameter at the cost of
being non-local, which is not a hindrance for a numerical study; a
MC sampling of the gauge fields weighted by \eqn{partition} becomes
simple in the Fourier basis where the Laplacian is diagonalized and
the modes are decoupled. We absorbed the fundamental real-valued
charge $q$ in \eqn{cqed} in a redefinition of gauge fields when
defining the parameterless lattice model, and hence the observables
will couple to gauge fields as $q\theta_\mu(x)$, or integer multiples
thereof.  We discuss further details of the model and the algorithm
in the Supplementary Material, which includes
Refs.~\cite{Sulejmanpasic:2019ytl,Villain:1974ir,Pufu:2013vpa,Polyakov:1975rs,Karthik:2018rcg,Karthik:2019mrr}.

The absence of tunable parameters in the lattice action by itself
is not an indication of it being critical. A strong evidence of the
scale invariant behavior was seen in the sole dependence on
aspect-ratio $\zeta=l/t$ of all $l\times t $ Wilson loops, ${\cal
W}(q\theta)$, after a simple perimeter term is removed. The asymptotic
behavior~\cite{Peskin:1980ay} is characterized by $\nu \zeta$ as
$\zeta\to\infty$ and $\frac{\nu}{\zeta}$ for $\zeta\to 0$ with the
coefficient $\nu=-0.0820(8)q^2$ that should be universal for all
theories approaching this CFT, such as QED$_3$ (refer Supplementary
Material).  Another interesting pure-gauge observable is the
topological current,
$V^{\rm top}_\mu\equiv
\frac{q}{4\pi}\sum_{\nu\rho}\epsilon_{\mu\nu\rho}F_{\nu\rho}$, 
which
is trivially conserved in this noncompact U(1) theory.  We also
checked that its two point function for $1\ll |x|\ll L/2$ behaves
like a conserved vector correlator $\sum_{\mu} \left\langle V^{\rm
top}_\mu(0) V^{\rm top}_\mu(x)\right\rangle = C_V^{\rm top}|x|^{-4}$,
with the coefficient $C_V^{\rm top}=\frac{q^2}{4\pi^4}$ as expected
from the continuum regulated
calculation~\cite{Giombi:2016fct,Huh:2013vga,Chester:2016wrc}.  The
trivial $q^2$ dependence of conformal data in pure-gauge observables
becomes nontrivial in gauge invariant observables formed out of
spectator massless fermions.

\bef
\centering
\includegraphics[scale=0.65]{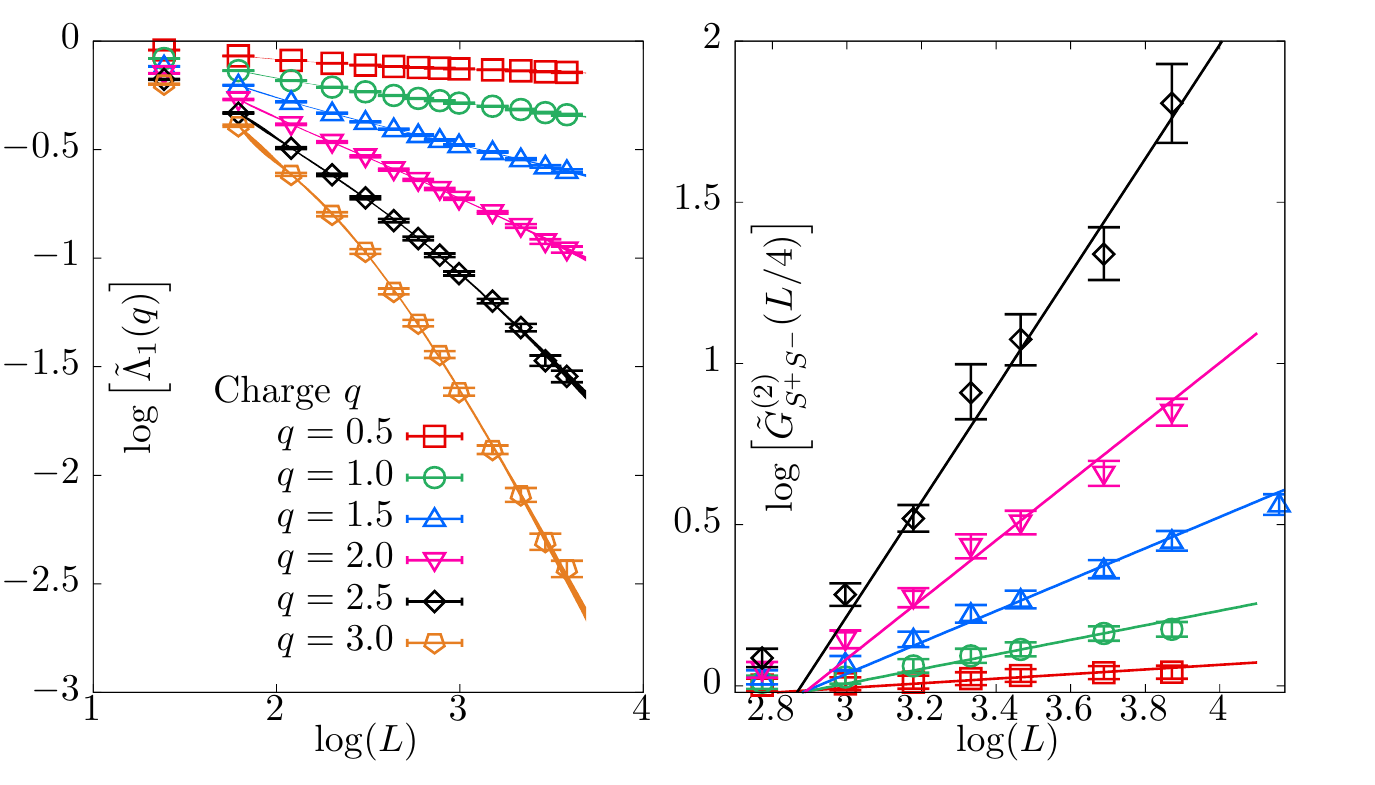}
\caption{Mass anomalous dimension as computed at different charges
$q$.  Left: The dependence of smallest Dirac eigenvalue $\tilde{\Lambda}_1(q)$, normalized by 
free theory value, on $L$. The curves are the fits to extract the leading $L^{-\gamma_S}$ dependence. 
Right: The finite size-scaling of the scalar two-point
function $\tilde{G}(|x_{12}|)$ at separations $|x_{12}|=L/4$. The
lines are the expected asymptotic dependence $\tilde{G}(|x_{12}|=L/4)\sim
L^{2\gamma_S}$ at different $q$, with $\gamma_S$ determined from $\tilde{\Lambda}_n$.
}
\eef{scalardim}

{\sl Conformal data in fermionic observables. --}
The lattice model per se does not have dynamical fermions. But, one
can couple spectator massless fermion sources to the model in order
to construct a variety of gauge-invariant hadronic correlation functions. 
Formally, the source term for a pair of parity-conjugate Dirac
fermions is $\bar{\psi}^+_q{\cal G}_q\psi^+_q-\bar{\psi}^-_q{\cal
G}_q\psi^-_q$, where ${\cal G}_q$ is the exactly massless overlap lattice
fermion propagator~\cite{Karthik:2016ppr,Kikukawa:1997qh,Narayanan:1994gw}
 coupled to the gauge-fields through the gauge-links
$e^{iq\theta_\mu(x)}$ (see Supplementary Material for the 
implementation of overlap Dirac operator, which includes Refs.~\cite{Chiu:2002eh,Hasenfratz:2001hp}). 
The flavor-triplet fermion bilinears are {\sl defined}
by taking appropriate derivatives
\beqa
&& O^\pm(x;q)=\left(\frac{\partial}{\partial\bar{\psi}^\pm_q}\Gamma\frac{\partial}{\partial \psi^\mp_q}\right)(x);\cr 
&& O^0(x;q)=\frac{1}{\sqrt{2}}\left(\frac{\partial}{\partial\bar{\psi}^+_q}\Gamma\frac{\partial}{\partial \psi^+_q} + \frac{\partial}{\partial\bar{\psi}^-_q}\Gamma\frac{\partial}{\partial \psi^-_q}\right)(x),
\eeqa{bilin}
of the effective action; $\Gamma=1$ for scalar bilinear, $S^{\pm,0}$,
and Pauli matrices $\Gamma=\sigma_\mu$ for the conserved vector
bilinears, $V_\mu^{\pm,0}$.  Practically, this procedure is
equivalent to a prescription of replacing fermion lines with massless
fermion propagators to form gauge-invariant observables.  We also
imposed anti-periodic boundary conditions on fermion sources in all
three directions which is symmetric under both lattice rotation and
charge conjugation while removing the issue of trivial Dirac zero
modes present even in the free field $q=0$ limit. We will denote
the $n$ point functions formed out of these fermion bilinears by
$G^{(n)}(x_{ij};q)$ and the dependence on the $x_{ij}$, the separation
between the location of the $i^{\rm th}$ and $j^{\rm th}$
bilinears should match the structure deduced from conformal symmetry.
Since we are only interested in changes to observables from free-field theory, we
form the ratios
$\tilde{G}^{(n)}(x_{ij};q)=G^{(n)}(x_{ij};q)/G^{(n)}(x_{ij};0)$,
which we henceforth refer to as {\sl reduced} $n$-point functions;
this also helps decrease any finite-size and short-distance lattice
effects that are already present in the free-field case.

We define scaling dimensions $\Delta_i=2-\gamma_i$ governing the
scaling $\tilde{G}^{(2)}_{O_i O_i}(x_{12})=C_i|x_{12}|^{2\gamma_i}$
for distances larger than few lattice spacings. The scaling dimension
$\Delta_S(q)=2-\gamma_S(q)$ of $S^{\pm,0}$ is an example of nontrivial
conformal data that is induced in this model. The $q$-dependent
non-zero $\gamma_S$ can be obtained from the finite-size scaling
(FSS) of the scalar two-point function, $\tilde{G}^{(2)}_{S^+S^-}(|x|=\rho
L)=L^{2\gamma_S} \left(g(\rho)+{\cal O}(1/L)\right)$ at fixed $\rho$.
The data for $\log\left[\tilde{G}^{(2)}_{S^+S^-}\right]$ at $\rho=1/4$
is shown as a function of $\log(L)$ using values of $q$ ranging
from $q=0.5$ to 2.5 in the right panel of \fgn{scalardim}, and one
sees that the slope of $\log(L)$ dependence (which is $2\gamma_S$)
increases monotonically from 0 when $q$ is increased.  Better
estimates of $\gamma_S(q)$ were obtained by studying the  FSS of
the low-lying discrete overlap-Dirac eigenvalues $\Lambda_j(L;q)$,
satisfying ${\cal G}^{-2}_q v_j=-\Lambda^{2}_j v_j$; the FSS,
$\Lambda_j(L;q)\propto L^{-1-\gamma_S(q)},$ is a consequence of the
FSS of the scalar susceptibility. In the left panel of \fgn{scalardim},
we show the reduced eigenvalues, $\tilde{\Lambda}_j(L;q)\equiv
\Lambda_j(L;q)/\Lambda_j(L;0)$ for $j=1$ as a function of $L$ along
with curves from combined fits using a functional form
$\tilde{\Lambda}_j(L;q)=a_jL^{-\gamma_S}(1+\sum_k^4 b_{jk} L^{-k})$
to first five $\tilde{\Lambda}_j$  using data from $L=6$ up to
$L=36$ (refer Supplementary Material, which includes
Ref.~\cite{Kalkreuter:1995mm}). Such a functional form with leading
scaling behavior and subleading scaling corrections nicely describes
the data and leads to precise estimates of $\gamma_S(q)$ that
increases continuously from $\gamma_S=0$ to ${\cal O}(1)$ in the
vicinity of $q\approx 2$; this dependence is captured to a good
accuracy by $\gamma_S(q)=0.076(11) q^2+0.0117(15) q^4+{\cal O}(q^6)$,
over this entire range of $q$.  For some charge $q=q_c\approx 2.9$,
the value of $\gamma_S$ becomes greater than 1.5, which is the
unitarity bound on scalars in a three-dimensional CFTs
(c.f.,~\cite{Simmons-Duffin:2016gjk}); therefore, within the framework
of constructing fermionic observables in this pure gauge theory,
we need to restrict ourselves to values of $q<q_c$ to be consistent
with being an observable in a CFT.  Unlike the scalar bilinear,
$V^a_\mu$ is conserved current and hence, does not acquire an
anomalous dimension. Therefore, the only non-trivial conformal data
is the two-point function amplitude,
$C_V(q)=\sum_{\mu=1}^3\tilde{G}^{(2)}_{V^a_\mu V^a_\mu}(|x|;q)$
that we were able to obtain from the plateau in the reduced vector
two-point correlator as a function of separations, $0\ll |x|\ll
L/2$ (refer Supplementary Material).  Its $q$-dependence can be
parameterized as $4\pi^2 C_V(q)=1-0.0478(7)q^2+0.0011(2)q^4+{\cal
O}(q^6)$.

\bef
\centering
\includegraphics[scale=0.34]{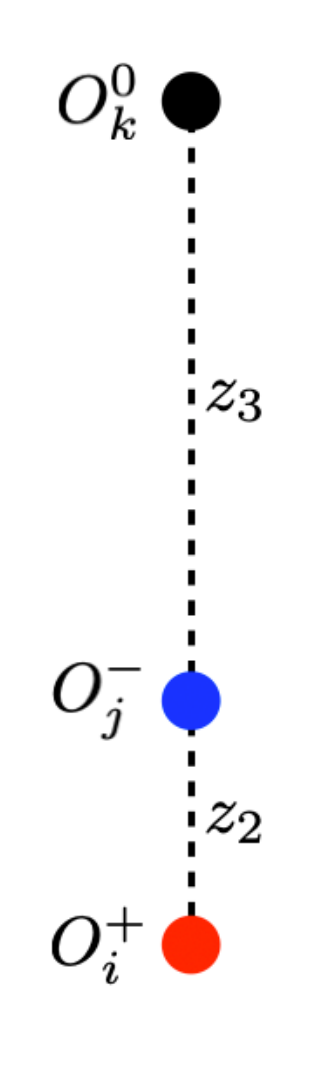}
\includegraphics[scale=0.61]{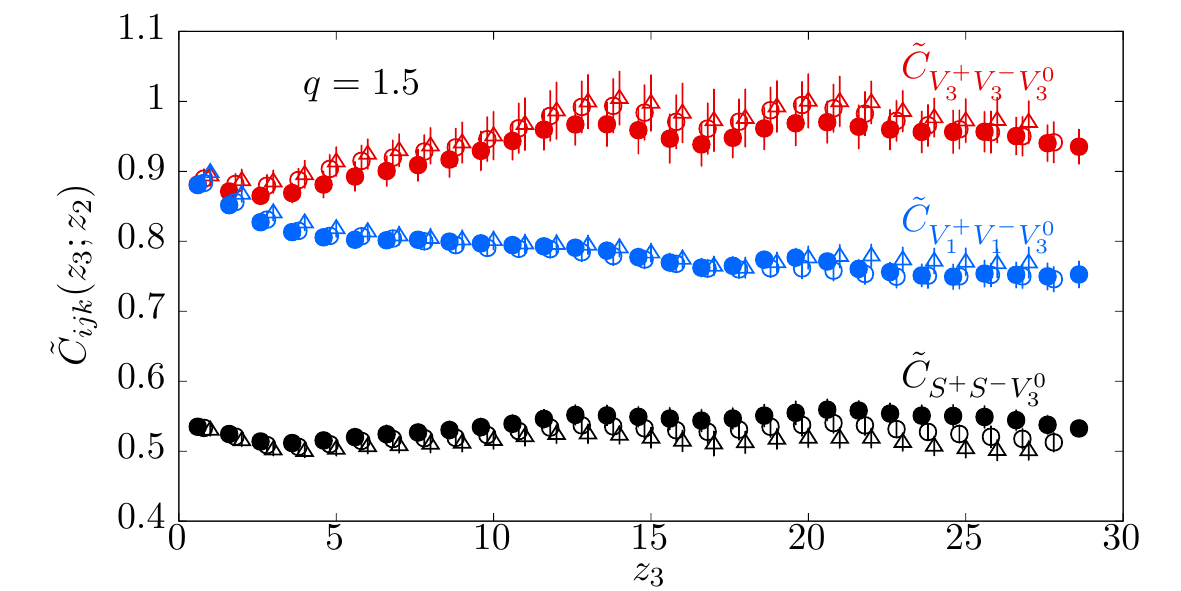}
\caption{Left: A configuration of collinearly placed operators.
Right: The effective OPE coefficients $\tilde{C}_{ijk}(z_2,z_3;q)$
of three different collinear three-point functions (distinguished by colors and 
slightly displaced)
are shown as a function of $z_3$ at three different fixed $z_2=$6
(open triangles), 8 (open circles), 10 (filled circles).  }
\eef{wilcoeff}

In order to demonstrate further the efficacy of the model as a CFT
with non-trivial conformal data in the massless spectator fermion
observables that is tractable numerically on the lattice, we also
present a proof-of-principle computation of the OPE coefficients
$\tilde{C}_{ijk}(q)$ of the reduced three-point functions
$\tilde{G}^{(3)}_{O_1O_2O_3}(x_{12},x_{23},x_{31};q)$ when three
operators lie collinearly, that is, $x_1=(0,0,0)$, $x_2=(0,0,z_2)$
and $x_3=(0,0,z_2+z_3)$ as described in the left panel of \fgn{wilcoeff}.
We looked at three distinct three-point functions, chosen so as to
reduce finite size effects, and whose dependences are fixed by
conformal invariance~\cite{Osborn:1993cr} to be
\beqa
\tilde{G}^{(3)}_{V^+_\mu
V^-_\mu V^0_3}(z_2,z_3)&=&\tilde{C}_{V^+_\mu V^-_\mu V^0_3};\ \ \mu=\mu_\perp(=1,2) \text{\ or\ } 3,\cr
\tilde{G}^{(3)}_{S^+ S^- V^0_3}(z_2,z_3) &=& \tilde{C}_{S^+ S^- V^0_3} z_2^{2\gamma_S},
\eeqa{tptfunc}
when $0\ll z_2,z_3,z_2+z_3 \ll L/2$ on a periodic lattice. For any
other separations, we use these expressions to define the effective
$z_2$ and $z_3$ dependent OPE coefficients which will display a
plateau as a function of $z_2,z_3$ provided the theory is a CFT.
In the right part of \fgn{wilcoeff}, we show 
the three effective OPE coefficients as a function of $z_3$ at three
different fixed $z_2(=6,8,10)$ as determined on $64^3$ lattice using
$q=1.5$. The plot demonstrates the independence of the three
coefficients on $z_3$ by a plateau over a wide range of $z_3$ that
is not too small or too large.  It also demonstrates their independence
on $z_2$ since the data from three different intermediate values
of $z_2$ are consistent, with this being quite non-trivial especially
for $\tilde C_{S^+S^-V_3^0}$ as it comes from a cancellation with a factor
$z_2^{2\gamma_S}$.  The conformal symmetry in general allows
non-degenerate OPE coefficients $\tilde{C}_{V^+_3 V^-_3 V^0_3} = \frac{a+b}{b_0}$
and $\tilde C_{V^+_{\mu_\perp} V^-_{\mu_\perp} V^0_3}=\frac{b}{b_0}$, with $a=0, b=b_0$
in free theory.  From \fgn{wilcoeff}, it is evident that $a\ne 0$
and $b\ne b_0$, clearly indicating that the result is for an
interacting CFT.
\bef
\centering
\includegraphics[scale=0.6]{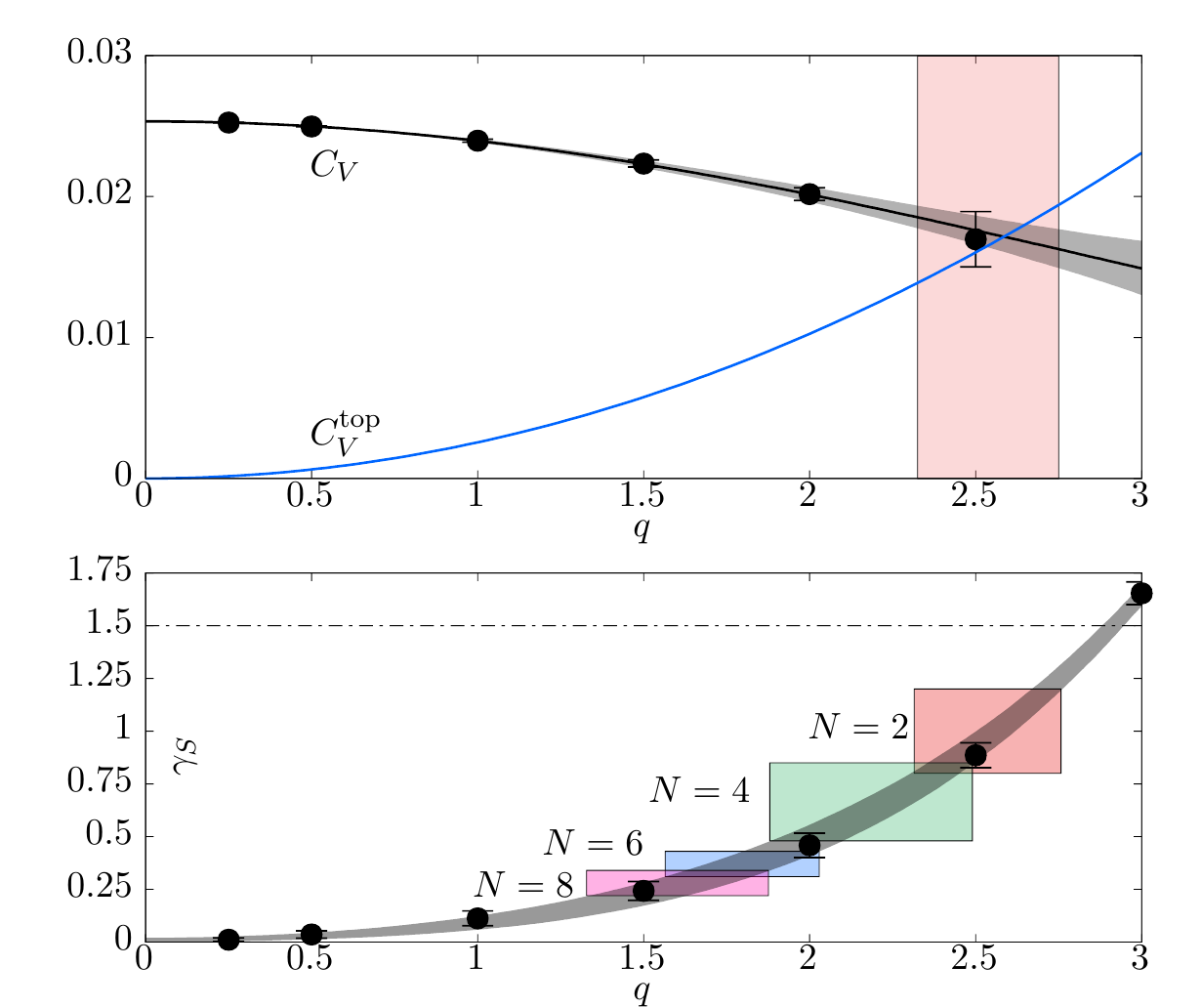}
\caption{
Bottom panel: 
Mass anomalous dimension $\gamma_S$ is shown as a function of charge $q$.
The filled circles are numerical determinations in the lattice model and the black band 
is the resulting spline interpolation of the data. 
The expected region corresponding to $N=2,4,6,8$ flavor 
QED$_3$ are shown by the rectangular boxes, so as to match the values of $\gamma_S$. 
The dashed line is the unitarity bound on $\gamma_S$.
Top panel: The $C_V$ in the lattice model
and $C_V^{\rm top}=q^2/(4\pi^4)$ are shown as a function of $q$. The two
intersect in the region of $q$ corresponding to $N=2$ QED$_3$, as inferred from the bottom panel.
}
\eef{match}

{\sl Relevance of the model to QED$_3$. --}
We will show a correspondence between the behavior of the CFT
at one particular $q$ and QED${}_3$ with $N$ flavors of massless
two component fermions.  Our surprising observation for
which we will present empirical evidences is that, for any finite
$N$, as long as QED$_3$ flows to an infrared fixed point, the
dominant effect of fermion determinant in QED$_3$ path-integral is to
induce a non-local quadratic conformal action for the gauge fields with a coupling
$q={\cal Q}(N)$ for some function ${\cal Q}$ that has to be determined
{\sl ab initio}, with the only condition being ${\cal Q}(N)\sim
\sqrt{32/N}$ for {\sl large} values of $N$.  That is, if the map
${\cal Q}(N)$ is known for all $N$, then one can study universal
features of the $N$-flavor QED$_3$ by studying the same properties
in the conformal lattice model at the corresponding $q={\cal Q}(N)$
with non-dynamical massless fermion sources, whose purpose is simply
to aid the construction of fermionic $n$-point functions. In order
to find ${\cal Q}(N)$, we propose to map values of $q$ in the
lattice model to $N$ in QED$_3$ such that the values of scalar
anomalous dimensions $\gamma_S$, determined non-perturbatively in
both theories, are the same. Such an identification of $q$ and $N$
is made in the bottom panel of \fgn{match}, where we have plotted
$\gamma_S(q)$ as a function of $q$, and determined expected 1-$\sigma$
ranges of $q$ that corresponding to $N=2,4,6,8$ flavor QED$_3$ based
on estimates of $\gamma_S$ from our previous lattice studies of
QED$_3$~\cite{Karthik:2015sgq,Karthik:2016ppr} ; namely, we find
the expected ranges $q\in[2.32,2.76], [1.88,2.49], [1.57,2.03], [1.33,1.88]$
for $N=2,4,6,8$ respectively. Below, we discuss two consequences of this connection.

\bef
\centering
\includegraphics[scale=0.5]{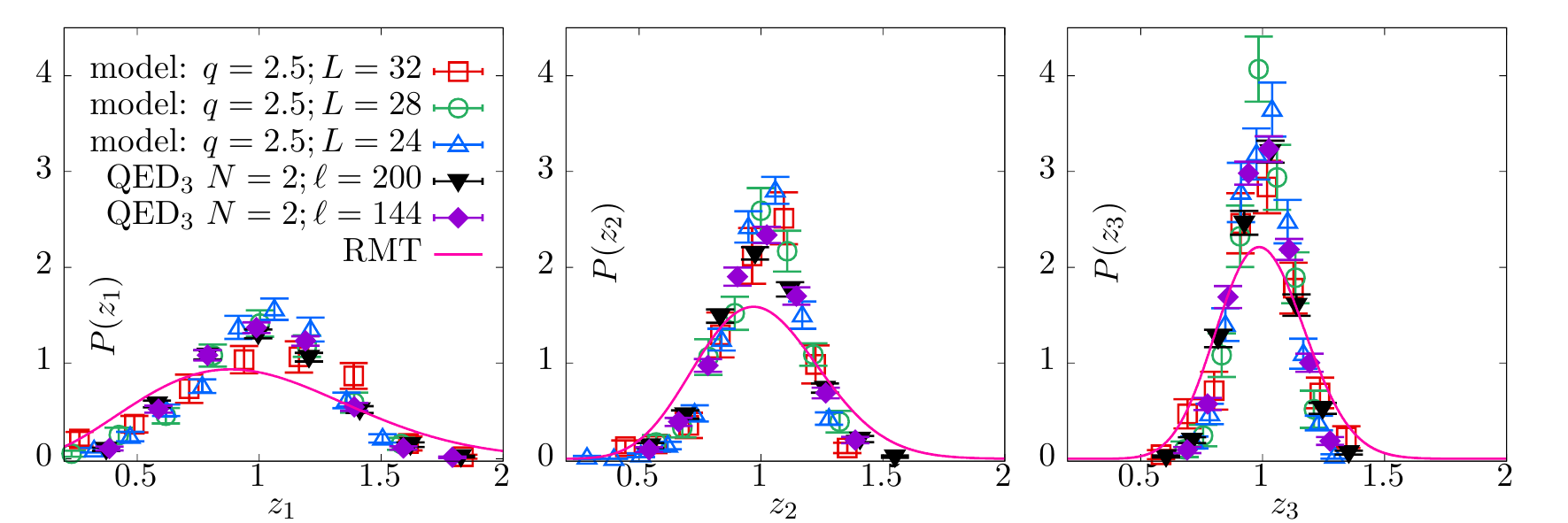}

\includegraphics[scale=0.5]{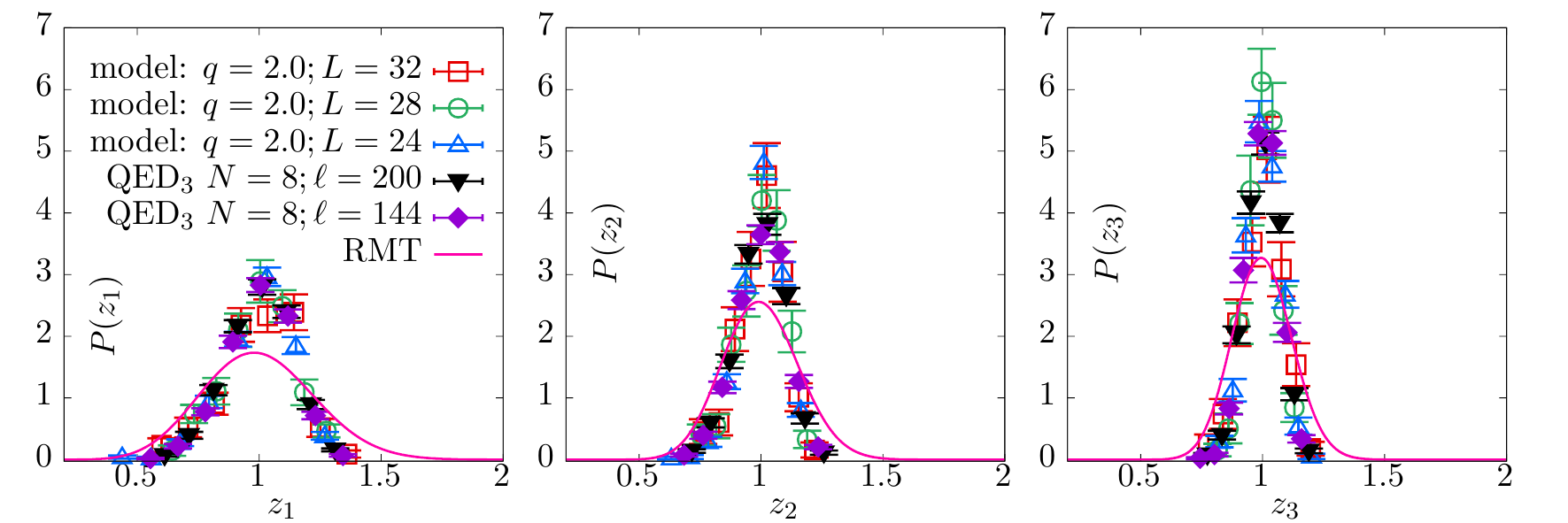}
\caption{
    Distribution of scaled eigenvalues $z_i=\frac{\Lambda_i}{\langle\Lambda_i\rangle}$ 
    for the three lowest eigenvalues (left to right) from the conformal lattice model
    at $q=2.5$ (top) and $q=2.0$ (bottom) are compared with those from
    $N=2$ and $N=8$ QED$_3$. For the lattice model, results from $L=24,28,32$ are shown,
    where as for QED$_3$, results from two large box sizes $\ell$ (measured in units of 
    coupling $g^2$) are shown. 
}
\eef{histo}

In the lattice model, the two-point functions of both $V^a_\mu$ and
$V_\mu^{\rm top}$ behave as $|x|^{-4}$ with amplitudes $C_V(q)$
having a non-trivial dependence on $q$ and $C_V^{\rm top}(q)$ being
quadratic in $q$. In the top-panel of \fgn{match}, we have shown
these $q$-dependences of the two amplitudes, wherein one finds
$C_V^{\rm top}$ increases as $q^2/(4\pi^4)$ whereas $C_V$ decreases
from the free field value $1/(4\pi^2)$ as a function of $q$, and
the two curves intersect around $q=2.6$; at this intersecting point,
$(V^+_\mu,V^0_\mu,V^-_\mu,V^{\rm top}_\mu)$ form an enlarged set
of degenerate conserved vector currents in the lattice model. It
is fascinating that this value of $q\approx 2.6$ lies in the probable
range corresponding to $N=2$ QED$_3$, where such a degeneracy is
expected from a conjectured self-duality of $N=2$
QED$_3$~\cite{Wang:2017txt,Xu:2015lxa,Hsin:2016blu} (conditional
to the theory being conformal), and the $q$-$N$ mapping presented
here suggests that such a degeneracy could occur in $N=2$ QED$_3$
(and also numerically observed in~\cite{Karthik:2017hol}).

Quite strikingly, we also find evidence for microscopic matching
between QED${}_3$ and the conformal model studied in this paper.
The probability distribution $P(z_i)$ of the scaled low-lying
discrete Dirac eigenvalues $z_i=\Lambda_i/\langle\Lambda_i\rangle$
are universal to QED$_3$ in the infrared limit and the lattice model
at the matched point ${\cal Q}(N)$.  In the top panels of \fgn{histo},
we show the nice agreement between $P(z_i)$ for the lowest three
eigenvalues from $N=2$ QED$_3$ at two different large box sizes
$\ell$ (measured in units of Maxwell coupling $g^2$)~\cite{Karthik:2015sgq,Karthik:2016ppr}
which are in the infrared regime, and the distributions $P(z_i)$
from the lattice model discussed here at $q=2.5$ which lies in the expected range
of $q$ for $N=2$.  Such an agreement is again seen between $P(z_i)$
in the lattice model at $q=2.0$ (which lies near the upper edge of
the expected range of $q$ for $N=8$) and in $N=8$ QED$_3$ shown in
the bottom panels. To contrast, such universality in low-lying
eigenvalue distribution has previously been studied only between
fermionic theories with a condensate and random matrix theories
(RMT) with same global symmetries~\cite{Verbaarschot:1994ip}. The
results for $P(z_i)$ from non-chiral RMT~\cite{Verbaarschot:1994ip}
corresponding to $N=2$ and 8 flavor theories are also shown for
comparison in top and bottom panels of \fgn{histo}, using analytical 
results in~\cite{Nishigaki:2016nka,Damgaard:1997pw}; the observed
disagreement between $P(z_i)$ in $N\ge 2$ QED$_3$ and the corresponding
RMTs is an evidence for the absence of condensate in parity-invariant
QED$_3$ with any non-zero number of massless fermions (as previously
observed by us in~\cite{Karthik:2015sgq}), and instead, the striking
compatibility of the QED$_3$ distributions with those from a CFT
studied here is a remarkable counterpoint.

{\sl Discussion. --}
We have presented a three dimensional interacting conformal field
theory where one can compute conformal data by a lattice regularization
without fine tuning.  We showed that by probing this CFT with
massless spectator fermions, one is able to obtain a more elaborate
set of conformal data that is tunable based on the charge of the
fermions.  For the sake of demonstration, we only
computed two and three point functions of fermion bilinear that
have the same charge. A simple extension for the near future is a computation of 
$n$-point functions of four-fermi operators
$\bar{\psi}_{n_1 q}\bar{\psi}_{n_2 q}\psi_{n_3 q}\psi_{(n_1+n_2-n_3)q}$
that is gauge-invariant nontrivially and has only connected diagrams.
We demonstrated a direct correspondence between the model with
charge-$q$ fermions and an $N$-flavor QED$_3$; by tuning $q$ so as
to match a scaling exponent (we chose $\gamma_S$), one is able to
observe many other universal features between the two corresponding
theories.  We stress that we did not perform an all-order calculation
in $1/N$ for QED$_3$~\cite{Gracey:1993sn,Giombi:2016fct,Chester:2016ref}
via a lattice simulation of the model; rather, the lattice calculation
is an all-order computation in charge-$q$ which might or might-not
be expandable in $1/N$ via a mapping $q={\cal Q}(N)$ that we
determined by a non-perturbative matching condition. 
However, a lattice perturbation theory approach to the
results presented here would be interesting.
It would also be interesting to use this model to test for robust predictions of infrared
fermion-fermion dualities~\cite{Seiberg:2016gmd,Karch:2016sxi} by
tuning the value of $q={\cal Q}(N)$ and adding required
level-$k$ lattice Chern-Simons term $\det[(1-{\cal G})/(1+{\cal
G})]^k$~\cite{Karthik:2018tnh}.

\acknowledgments
R.N. acknowledges partial support by the NSF under grant number
PHY-1913010.  N.K. acknowledges support by the U.S. Department of
Energy under contract No. DE-SC0012704.

\bibliography{ref.bib}

\newpage
\clearpage
\pagestyle{empty}
\onecolumngrid

\begin{center}
{\Large Supplementary Material}
\end{center}

\section{The general U(1) lattice model: noncompact and compact theories}
In this appendix, we write down a general U(1) gauge theory, of
which the non-compact model considered in this paper is a specific
case. To avoid confusion, the terminology {\sl compact} and {\sl
non-compact} in the lattice field theory language means that they
are U(1) theories with and without monopoles
respectively~\cite{Karthik:2015sgq,Sulejmanpasic:2019ytl}. The U(1)
model, that in general has monopole defects, can be defined using
a Villain-type~\cite{Villain:1974ir} action:
\beq
Z\equiv \left(\prod_{x,\mu}\int_{-\infty}^\infty d\theta_\mu(x)\right)\sum_{\{N\}} e^{-S_g(N)}
\eeq{model1}
where
\beq
    S_g(N)=\frac{1}{2}\sum_{\mu,\nu=1}^3\sum_{x,y} \left[F_{\mu\nu}(x)-\frac{2\pi}{q}N_{\mu\nu}(x)\right]\left[\Box^{-1/2}\right](x,y)\left[F_{\mu\nu}(y)-\frac{2\pi}{q}N_{\mu\nu}(y)\right];\quad F_{\mu\nu}(x)=\Delta_\mu\theta_\nu(x)-\Delta_\nu\theta_\mu(x),
\eeq{model}
for integer valued fluxes $N_{\mu\nu}$ defined over plaquettes, and
$q$ is the real valued dimensionless charge.  The theory has the
U(1) gauge symmetry $\theta_\mu(x)\to \theta_\mu(x)+\Delta_\mu
\chi(x)$ as well as a symmetry $\theta_\mu(x)\to
\theta_\mu(x)+\frac{2\pi}{q}m_\mu(x)$ for integers $m_\mu$.  Fermions
sources $\psi_{nq}$ in this model couple to $\theta$ via compact
link variables $e^{i n q\theta_\mu(x)}$. Monopoles of integer valued
magnetic charges $q_{\rm mon}$ at a cube at site $x$ is given by
\beq
\frac{1}{2}\sum_{\rho,\mu,\nu=1}^3 \epsilon_{\rho\mu\nu}\Delta_\rho N_{\mu\nu}(x)=2\pi q_{\rm mon}(x).
\eeq{monodef}
The non-compact U(1) theory is a specific case obtained by the
restriction that the number of monopoles at any site $x$ is zero,
i.e., $q_{\rm mon}(x)=0$. This gives the condition that the integer
valued fluxed $N_{\mu\nu}(x)$ be writable as a curl of integer
valued links:
\beq
N_{\mu\nu}(x)=\Delta_\mu m_\nu(x)-\Delta_\nu m_\mu(x).
\eeq{divcond}
Under such a condition, the explicitly U(1) symmetric partition
function in \eqn{model1} can be equivalently written as the non-compact
action we study in this paper,
\beq
Z\equiv \left(\prod_{x,\mu}\int_{-\infty}^\infty d\theta_\mu(x)\right)  e^{-S_g(N=0)},
\eeq{ncaction}
by appropriately redefining $\theta_\mu(x)\to \theta_\mu(x)-\frac{2\pi}{q}
m_\mu(x)$ in the original action.  Such a connection also means
that the observables ${\cal O}(\theta)$ be restricted to those
invariant under $\theta_\mu(x)\to \theta_\mu(x)+\frac{2\pi}{q}m_\mu(x)$
for the equivalence of two ways of writing the U(1) theory without
monopoles. We only studied the non-compact action above in this
paper. 

A future study of the compact model with monopole degrees of freedom 
will be very interesting for the following reason.  In the weak-coupling limit of $q\to 0$, 
the monopoles will get suppressed energetically,
and hence be irrelevant, and we expect the theory would
remain conformal as the noncompact theory. This irrelevance of
monopoles might continue up to some critical $q=q_c$ beyond which
monopoles could become relevant (their scaling dimension become
smaller than 3)~\cite{Pufu:2013vpa}, and the theory could be confining
like the pure gauge compact Maxwell theory~\cite{Polyakov:1975rs}.  This
study will be feasible using the approaches presented
in~\cite{Karthik:2018rcg,Karthik:2019mrr}.

\section{Monte-Carlo algorithm in Fourier space}
The lattice action in real space is non-local, but it is diagonal
in momentum space.  In this appendix, we describe the Monte-Carlo
algorithm in momentum space to generate independent gauge field
configurations. Our convention for Fourier transform $\chi\to
\tilde\chi$ on the lattice is
\be
\chi(x) = {\sum}_{n_1,n_2,n_3=0}^{\prime L-1} \tilde \chi(n) e^{\frac{2\pi i n\cdot x}{L}};\qquad
\tilde \chi(n) = \frac{1}{L^3}{\sum}_{x_1,x_2,x_3=0}^{L-1} \tilde \chi(x) e^{-\frac{2\pi i n\cdot x}{L}}.
\ee
where
the prime over the sum denotes that the zero momentum mode $n=0$ is excluded.
The reality of a function $\chi(x)$ implies $\tilde \chi^*(n) = \tilde \chi(\bar n)$ where $\bar n_i = -n_i \ {\rm mod}\ L$.
with the lattice momentum given by $f_\mu(n) = e^{\frac{2\pi i n_\mu }{L}}-1$, the lattice action for the model in \eqn{partition} can be written as
\be
S_g 
= \frac{ L^3}{2} \sum_n \sum_{\mu ,\nu}^3 \frac{1}{\sqrt{f^{2}(n)}} \left [ f_\mu(\bar n) \tilde\theta_\nu(\bar n) - f_\nu(\bar n) \tilde\theta_\mu(\bar n) \right]  
\left [ f_\mu(n) \tilde\theta_\nu(n) - f_\nu(n) \tilde\theta_\mu(n) \right];\qquad f^2(n) \equiv \sum_{\mu=1}^3 f_\mu(n) f^*_\mu(n).
\ee
Assuming we will only be interested in computing observables that are gauge invariant, we will generate the two physical degrees of freedom
per momentum that are perpendicular to the zero mode,
\be
\tilde\theta_\parallel = \begin{pmatrix} f_1(n) \cr f_2(n)\cr f_3(n) \cr \end{pmatrix}.
\ee
We are free to pick the two directions perpendicular to the zero mode due to the degeneracy in this plane.
When $(n_1 +n_2) \ne 0$, we choose the normalized eigenvectors
\bea
\tilde\theta_{\perp 1} &=& \frac{1}{\sqrt{f_1^*(n) f_1(n) + f_2^*(n) f_2(n)}}
\begin{pmatrix} f_2^*(n) \cr -f_1^*(n) \cr 0 \end{pmatrix};\cr
\tilde\theta_{\perp 2} &=& \frac{1}{f(n)\sqrt{f_1^*(n) f_1(n) + f_2^*(n) f_2(n)}}
\begin{pmatrix} f_3^*(n)f_1(n) \cr f_3^*(n) f_2(n) \cr -f_1^*(n)f_1(n)-f_2^*(n) f_2(n) \end{pmatrix},
\eea
and when $n_1=n_2=0$, we choose
\be
\tilde\theta_{\perp 1} = \begin{pmatrix} 1 \cr 0 \cr 0 \end{pmatrix};\qquad
\tilde\theta_{\perp 2} = \begin{pmatrix} 0 \cr 1 \cr 0 \end{pmatrix}.
\ee
With these choice, the Monte Carlo algorithm is simple;
\begin{enumerate}
    \item Pick random numbers $c_{1,\mu}(n),c_{2,\mu}(n)\sim {\cal N}\left[\mu=0,\sigma^2=1/(L^3\sqrt{f^2(n)})\right]$.
    \item Construct $\tilde{\theta}_\mu(n)=c_{1,\mu}(n)\tilde\theta_{\perp 1,\mu}(n)+c_{2,\mu}(n) \tilde\theta_{\perp 2,\mu}(n)$.
\item Construct the gauge fields in real space as $\theta_\mu(x)=\sum^\prime_n \tilde\theta_\mu(n)e^{\frac{2\pi i n\cdot x}{L}}$.
\end{enumerate}
Just as a similar algorithm for pure gauge Maxwell theory, the
Monte-Carlo algorithm for this conformal action is free of auto-correlation
by construction. The expense of the anti-Fourier transform in the
last step can be drastically reduced by using a standard Fast Fourier Transform algorithm.

\section{Topological current correlator}
The topological current is
\be
V^{\rm top}_\mu(x) = \frac{q}{4\pi}\sum_{\nu,\rho=1}^3 \epsilon_{\mu\nu\rho} F_{\nu\rho},
\ee
which is conserved on the lattice.
To compute the two-point function, the source for $V^{\rm top}_\mu(x)$ is added as
\be
S_J = \frac{q}{2\pi} \sum_{\mu\nu\rho} \sum_{x_1,x_2,x_3=0}^{L-1} \epsilon_{\mu\nu\rho} J_\mu (x) \Delta_\nu\theta_\rho(x) 
=\frac{q L^3}{4\pi }  \sum_{\mu\nu\rho} {\sum}_{n_1,n_2,n_3=0}^{' L-1} \tilde\theta^*_\mu(n) \left( \epsilon_{\mu\nu\rho} f_\nu^*(n) \tilde J_\rho(n)\right)+ {\rm cc},
\ee
and only couples to $\tilde\theta_{\perp j}$ as expected. Then 
\be
\ln \frac{Z(k,J)}{Z(0)} = \frac{1}{2} \sum_{x,y} J_\mu(x) G_{\mu\nu}(x-y) J_\nu(y),
\ee
where
\be
G_{\mu\nu}(x) = \langle V_\mu^{\rm top}(x) V^{\rm top}_\nu(0)\rangle=\frac{q^2}{8\pi^2 L^3} {\sum}_{n_1,n_2,n_3=0}^{\prime L-1} \frac{\delta_{\mu\nu} f^2(n) - f^*_\mu f_\nu}{\sqrt{f^{2}(n)}} e^{\frac{2\pi i n \cdot x}{L}}.
\ee
The two-point function traced over the directions becomes
\be
G^{(2)}_{V^{\rm top}}(x) = \sum_{\mu=1}^3 G_{\mu\mu}(x) = \frac{q^2}{4\pi^2 L^3} {\sum}_{n_1,n_2,n_3=0}^{\prime L-1} \sqrt{f^{2}(n)} e^{\frac{2\pi i n \cdot x}{L}}\xrightarrow{L\to\infty} \frac{q^2}{64\pi^4} \int_{-\pi}^\pi d^3 p \frac{e^{ip\cdot x}}{\sqrt{\sum_\mu \sin^2 \frac{p_\mu}{2}}}. 
\ee

\bef
\centering
\includegraphics[scale=0.7]{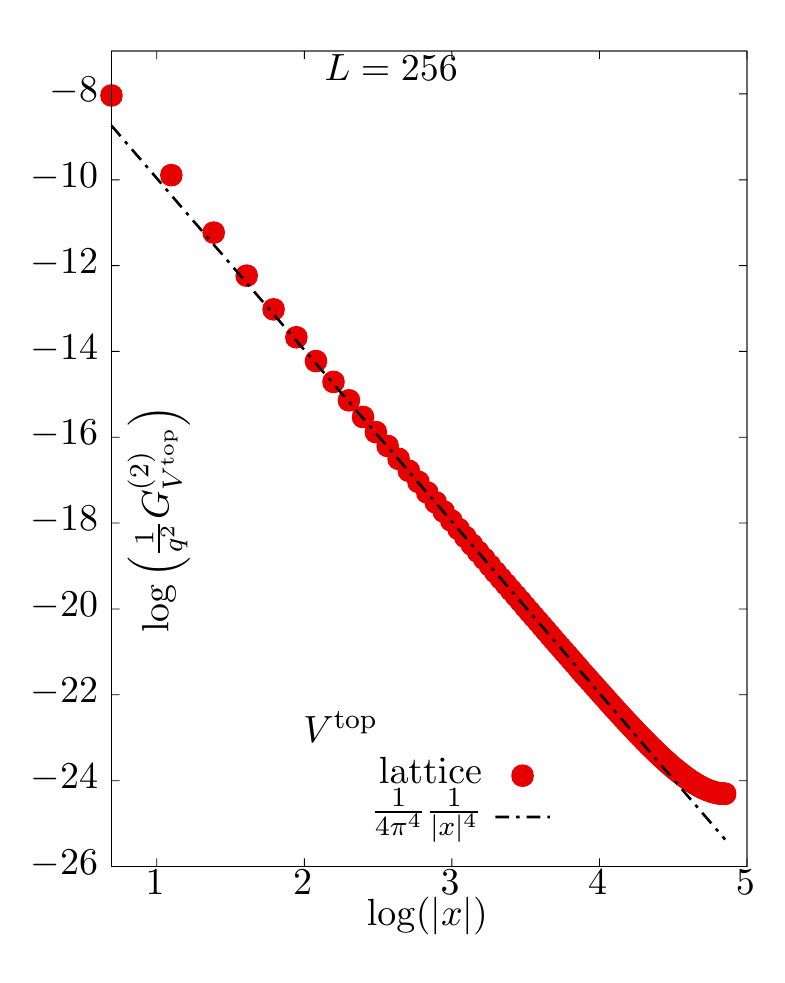}
\caption{
Topological current correlator, scaled by $1/q^2$, is plotted as a function of current-current 
separation $|x|$. The data is compared with the continuum expectation $1/\left(4\pi^4 |x|^4\right)$.
}
\eef{topcurrent}

In \fgn{topcurrent}, we plot $q^{-2} G^{(2)}_{V^{\rm top}}(x)$ as a function
of $|x|$ for $x=(0,0,z)$ as determined using the above expression
on $L=256$ lattice to show the effect of lattice regularization. For comparison, the continuum
result~\cite{Giombi:2016fct,Huh:2013vga,Chester:2016wrc} $q^{-2}
G^{(2)}_{V^{\rm top}}(x) = \frac{1}{4\pi^4}\frac{1}{|x|^4}$ is also plotted
as the black line.  It is clear for intermediate $1 \ll |x| \ll
L/2$, the value of $C_V^{\rm top}=q^2/(4\pi^4)$ is reproduced by
the lattice regularization. This intermediate range of $|x|$ indeed increases as
one keeps increasing $L$.

\section{Wilson-loop}
We consider $l\times t$ rectangular Wilson loop defined as 
\beq
{\cal W}_q(l,t)\equiv -\log\left\langle \exp\left(i
q\sum_{x\in l\times t} F_{\mu\nu}(x)\right)\right\rangle.
\eeq{wloopdef}
We compute its expectation value by 
coupling a source
\bea
J_1(x) &=& iq
 \sum_{y_1=0}^{l-1} \left[ \delta(x_1,z_1+y_1),\delta(x_2,z_2)\delta(x_3,z_3) - \delta(x_1,z_1+y_1),\delta(x_2,z_2+t)\delta(x_3,z_3) \right],\cr
J_2(x) &=& iq
 \sum_{y_2=0}^{t-1} \left[ \delta(x_1,z_1+l),\delta(x_2,z_2+y_2)\delta(x_3,z_3) - \delta(x_1,z_1),\delta(x_2,z_2+y_2)\delta(x_3,z_3) \right],\cr
J_3(x) &=& 0,
\eea
where $q$ denotes the charge.
Upon a Fourier transform the non-zero vectors are,
\be
\tilde J(n) = \begin{cases}
\frac{iq}{L^3} e^{-\frac{2\pi i n\cdot z}{L}} \left(1-e^{-\frac{2\pi i n_1l}{L}}\right) \left(1-e^{-\frac{2\pi i n_2t}{L}}\right)
\begin{pmatrix} -\frac{1}{f^*_1(n)} \cr \frac{1}{f^*_2(n)} \cr 0 \end{pmatrix}, & n_1, n_2 \ne 0;\cr
\frac{iq}{L^3} e^{-\frac{2\pi i n\cdot z}{L}}  \left(1-e^{-\frac{2\pi i n_2t}{L}}\right)
\begin{pmatrix} l \cr 0 \cr 0 \end{pmatrix}, & n_1=0, n_2 \ne 0;\cr
\frac{iq}{L^3} e^{-\frac{2\pi i n\cdot z}{L}} \left(1-e^{-\frac{2\pi i n_1l}{L}}\right) 
\begin{pmatrix} 0 \cr -t \cr 0 \end{pmatrix}, & n_1\ne 0, n_2 = 0;
\end{cases}
\ee
and we note that $\theta^\dagger_\parallel \tilde J(n) =0$, implying
that the Wilson loop operator only couples to the physical degrees
of freedom. The logarithm of the expectation value of the Wilson loop is proportional to $q^2$ and its
expression after factoring out the $q^2$  is
\bea
{\cal W}(l,t) &=& \frac{4}{L^3} {\sum}_{n_1=1,n_2=1,n_3=0}^{\prime L-1} \frac{1}{\sqrt{f^{2}(n)}} \sin^2 \frac{\pi n_1 l}{L} \sin^2 \frac{\pi n_2 t}{L} 
\left[\frac{1}{|f_1|^2} + \frac{1}{|f_2|^2}\right]\cr
&& +\frac{1}{L^3} {\sum}_{n_2=1,n_3=0}^{\prime L-1} \frac{l^2}{\sqrt{f^2(n;n_1=0)}} \sin^2 \frac{\pi n_2 t}{L} 
\cr
&& +\frac{1}{L^3} {\sum}_{n_1=1,n_3=0}^{\prime L-1} \frac{t^2}{\sqrt{f^2(n;n_2=0)}} \sin^2 \frac{\pi n_1 l}{L}\label{lwilson}
\eea
In the limit $L\to\infty$, we can write the above expression as an integral
\be
{\cal W}(l,t) = \frac{1}{2\pi^3}\int_{-\pi}^\pi d^3 p \frac{ \sin^2 \frac{p_1 l}{2} \sin^2\frac{p_2t}{2}}{\sqrt{\sum_{k=1}^3 \sin^2 \frac{p_k}{2}}}\left[ \frac{1}{\sin^2 \frac{p_1}{2}} + \frac{1}{\sin^2 \frac{p_2}{2}}\right].\label{cwilson}
\ee 

\subsection{Conformal behavior of Wilson loop}

\bef
\centering
\includegraphics[scale=0.7]{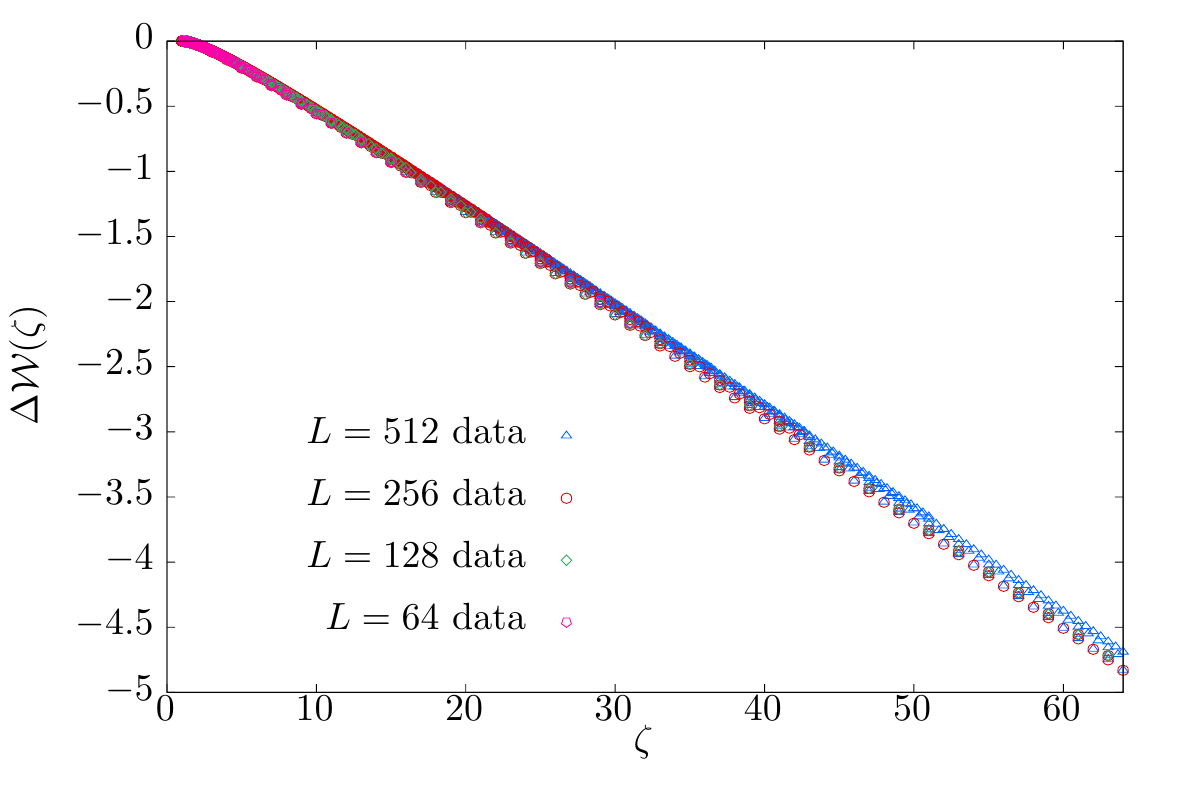}
\caption{
The data for the perimeter subtracted Wilson-loop $\Delta{\cal
W}(l,t)$ from multiple $(l,t)$ have been plotted together as a
function of $\zeta=l/t$. 
The near data collapse shows the dependence
only on $\zeta$. 
}
\eef{wloop}

\bef
\centering
\includegraphics[scale=0.7]{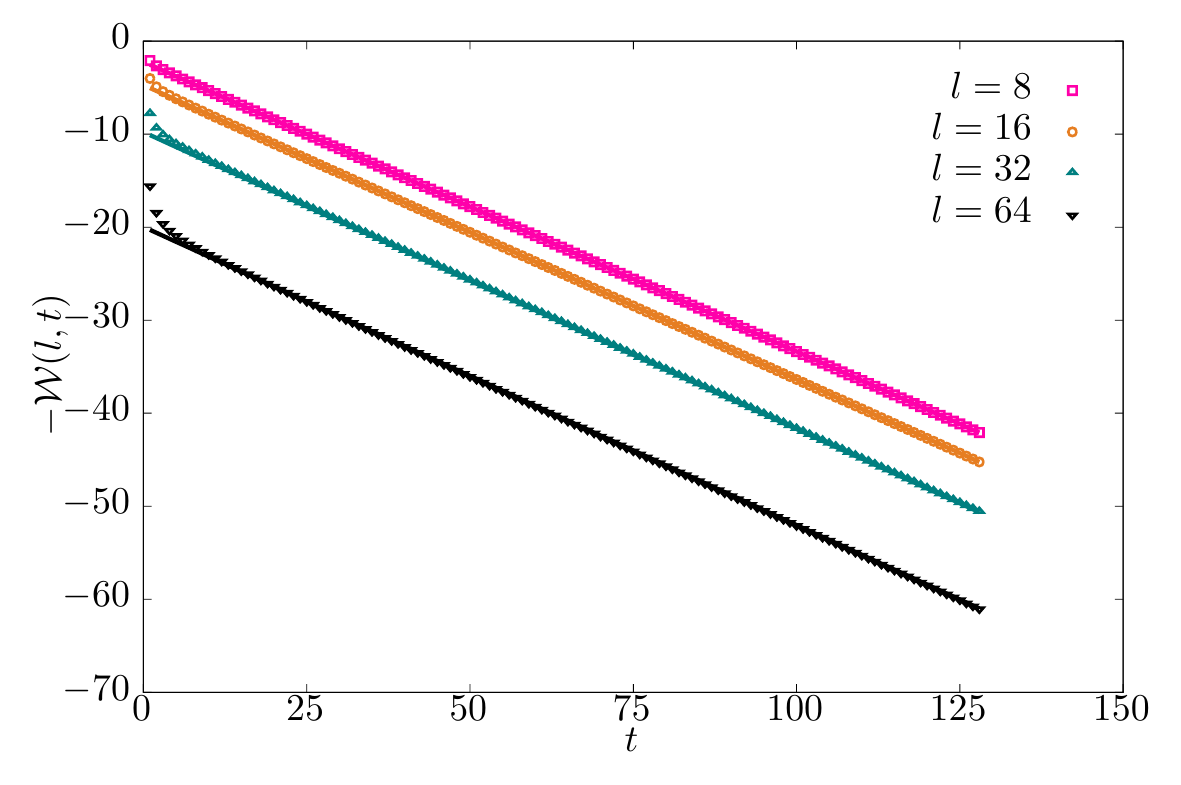}
\includegraphics[scale=0.7]{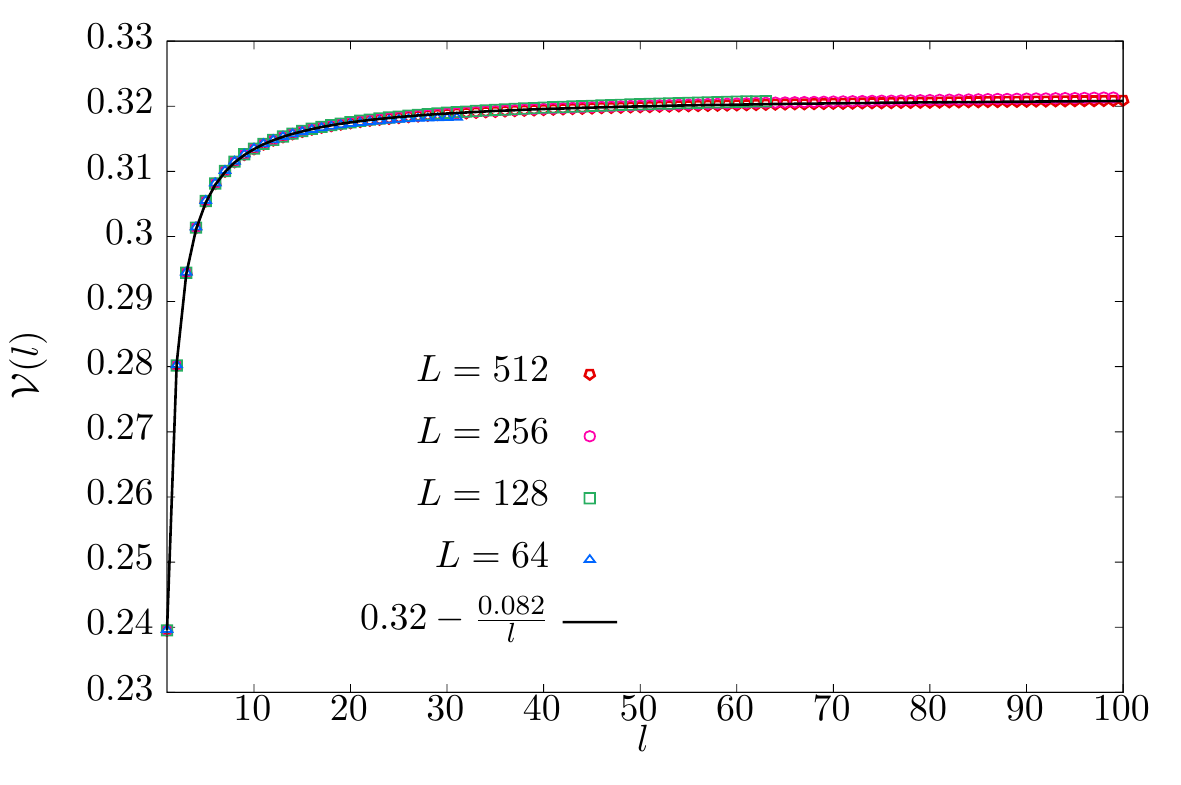}
\caption{
Extraction of static quark potential ${\cal V}(l)=\lim_{t\to\infty}
t^{-1} {\cal W}(l,t)$.  Left panel shows $-{\cal W}(l,t)$ as a
function of $t$ at different fixed $l$ on $L=256$ lattice. The
straight lines are fits to $-{\cal W}(t;l)=-A(l)-t {\cal V}(l)$
starting from $t=30$. The right panel shows the extracted static
fermion potential ${\cal V}(l)$ as function of $l$, and data from
different $L$ are shown. The data for $l< L/2$ can be well described
by a potential ${\cal V}(l)=0.322-0.0820/l$.  The $1/l$ is expected
in conformal gauge theories in any dimension.
}
\eef{pot}

The integral in \eqn{cwilson} results in a non-trivial dependence on $l$ and $t$ which includes a perimeter term.
We show that it is possible to extract the conformal behavior by evaluating the lattice sum in \eqn{lwilson}.
The semi-analytic expression above by itself is hard to understand;
hence we numerically evaluated the expressions for different $L^3$
lattices to determine the behavior of $l\times t$ rectangular Wilson
loops as a function of $l,t$.  Since the Wilson line depends on
charge as a simple $q^2$, we divide the results by $1/q^2$ and
present the results here (we will drop the index $q$ below.) In a
gauge theory which is critical, one expects ${\cal W}(l,t)$ to
depend only on the aspect ratio of the loop $\zeta=\frac{l}{t}$ up
to linear corrections from the perimeter of the loop, $p=l+t$.  In
\fgn{wloop}, we show the $\zeta$-dependence for the difference
\beq
\Delta{\cal W}(l,t)\equiv {\cal W}(l,t)-{\cal W}(p/2,p/2),
\eeq{deltaweq}
constructed such that any perimeter term gets canceled. For a given
$L$, Wilson loops of various possible $l$ and $t$ have been put
together in the plot. We have shown the results using $L=64,128,256$
and 512. One can see that the results from various $l\times t$ loops
fall on a universal curve to a good accuracy that depends only on
$\zeta$. This clearly demonstrates the underlying gauge theory is
conformal. At a fixed $\zeta$, one sees a little scatter of points
around a central value; this is because the lattice corrections
increase when the size of a Wilson-loop at a given $\zeta$ is
comparable to the lattice spacing itself. This can be justified by
observing that as $L$ is increased towards 512, the scatter of
points at given $\zeta$ becomes lesser, due to the possibility of
having larger loops with the same $\zeta$.  For large $\zeta$, one
finds a linear tendency of $\Delta{\cal W}(\zeta)$ originating from
the $1/t$ static potential as we discuss below.

We extract the static fermion potential ${\cal V}(l)$ 
by looking for the asymptotic behavior
\beq
{\cal W}(l,t)=A(l)+t {\cal V}(l)
\eeq{fitpot}
for larger $t$ at fixed $l$. For this, we fitted the above form to
${\cal W}(l,t)$ for $25 < t < L/2-10$, and obtained ${\cal V}(l)$,
using $L=64,128,256$ and 512. This is demonstrated in the left panel
of \fgn{pot} where $-{\cal W}$ is plotted as a function of $t$ for different fixed 
$l=8,16,32,64$ on $L=256$ lattice. The fits to the
above form are the straight lines. In the right panel of \fgn{pot},
we plot the extracted potential ${\cal V}(l)$ as a function of $l$.
We have shown the potential as extracted from $L=64,128$ and 256
as the different colored symbols. For $1< l \ll L/2$, the data is
nicely described by the form
\beq
{\cal V}(l)=0.322-\frac{0.0820}{l}.
\eeq{confpot}
It is important to remember that this functional form is not the
Coulomb potential in three dimensions (which is instead logarithmic
in 3d), and instead, this functional form is dictated by the conformal
invariance in gauge theories~\cite{Peskin:1980ay}.  The coefficient
$\nu\approx 0.0820$ is universal to theories approaching this CFT
(if one puts back the trivial charge $q$ dependence, for Wilson
loop of charge $q$, the coefficient will be $\nu(q)=0.0820 q^2$.)
By changing fit ranges, we find about 1\% variation in our estimates
of $\nu$; Therefore, we quote an estimate with a systematic
uncertainty, $\nu(q)=0.0820(8) q^2$.

\section{Overlap fermion propagator}
The details on the overlap formalism in three dimensions to study
exactly massless fermions on the lattice can be found
in~\cite{Karthik:2016ppr}. Here, we recall the important aspects
of the implementation of the overlap Dirac operator. The massless
overlap propagator ${\cal G}_q$ for a two-component Dirac fermion
of charge $q$ is given by
\beq
{\cal G}_q= \frac{V(q\theta)-1}{V(q\theta)+1},
\eeq{prop}
where $V(q\theta)$ is a unitary $2L^3\times 2L^3$ matrix. The matrix
$V$ is constructed using Wilson-Dirac operator kernel as
\beq
V(q\theta)\equiv \frac{1}{\sqrt{X(q\theta) X^\dagger(q\theta)}} X(q\theta).
\eeq{vdef}
$X$ is the Wilson-Dirac operator with mass $-m_w$,
\beq
X(q\theta)= \slashed{D}(q\theta)+B(q\theta)-m_w,
\eeq{xdef}
where $\slashed{D}$ and $B$ are the naive lattice Dirac operator
and the Wilson mass term respectively,
\beq
\slashed{D}(q\theta)=\frac{1}{2}\sum_{\mu=1}^3 \sigma_\mu\left[T_\mu(q\theta)-T^\dagger_\mu(q\theta)\right],\quad
B=\frac{1}{2} \sum_{\mu=1}^3\left[2-T_\mu(q\theta)-T_\mu^\dagger(q\theta)\right],
\eeq{ddef}
in terms of the covariant forward shift operator, $[T_\mu f](x)=e^{i
q \theta_\mu(x)} f(x+\hat\mu)$. The three Pauli matrices are denoted
as $\sigma_\mu$.

We improved the overlap operator by using 1-HYP
smeared~\cite{Hasenfratz:2001hp,Karthik:2015sgq} fields $\theta^s_\mu(x)$
instead of $\theta_\mu(x)$ in the above construction, which suppresses
gauge field fluctuations of the order of lattice spacing and in
particular, reduces the number of few lattice-spacing separated
monopole-antimonopole pairs which are artifacts in a noncompact
theory~\cite{Karthik:2016ppr}.  We implemented $(X X^\dagger)^{-1/2}$
by using Zolotarev expansion~\cite{Chiu:2002eh} up to 21st order,
which was found sufficient in~\cite{Karthik:2016ppr}. We used $m_w=1$
in the Wilson-Dirac kernel.

\section{Extraction of mass anomalous dimension from Dirac eigenvalues}
\befs
\centering
\includegraphics[scale=0.7]{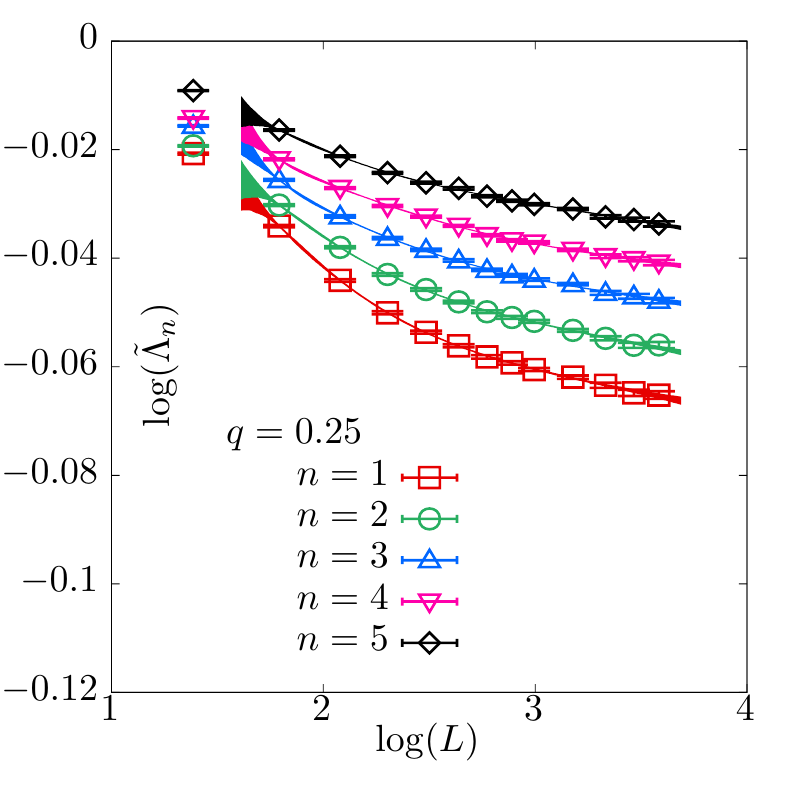}
\includegraphics[scale=0.7]{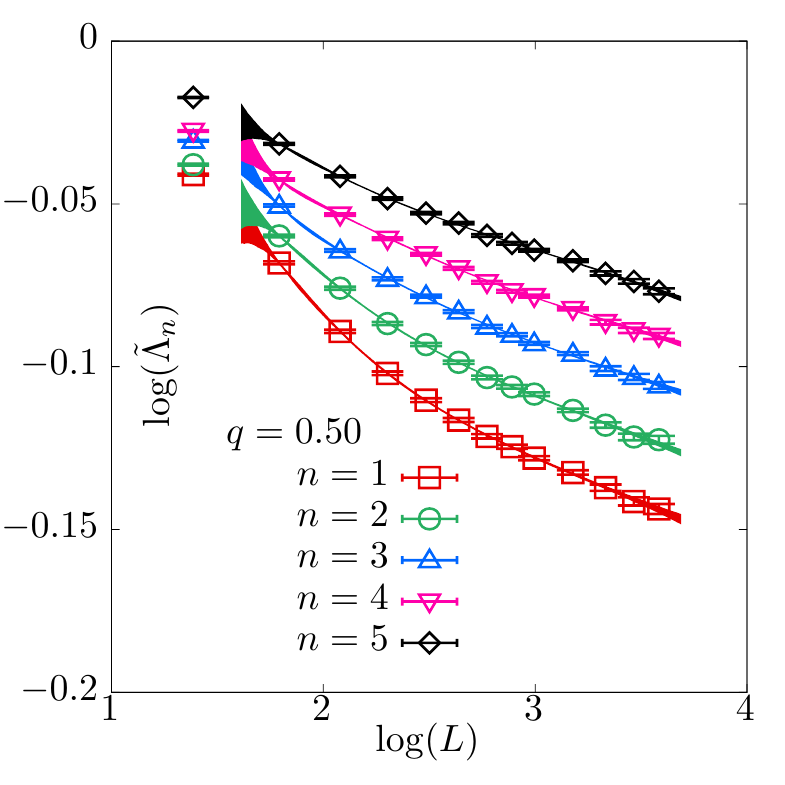}
\includegraphics[scale=0.7]{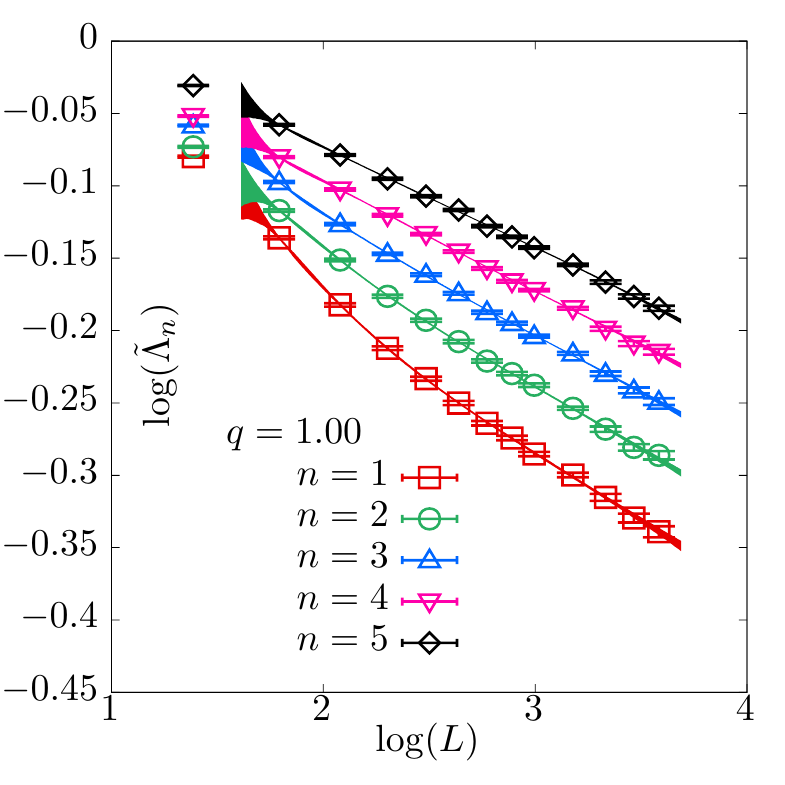}

\includegraphics[scale=0.7]{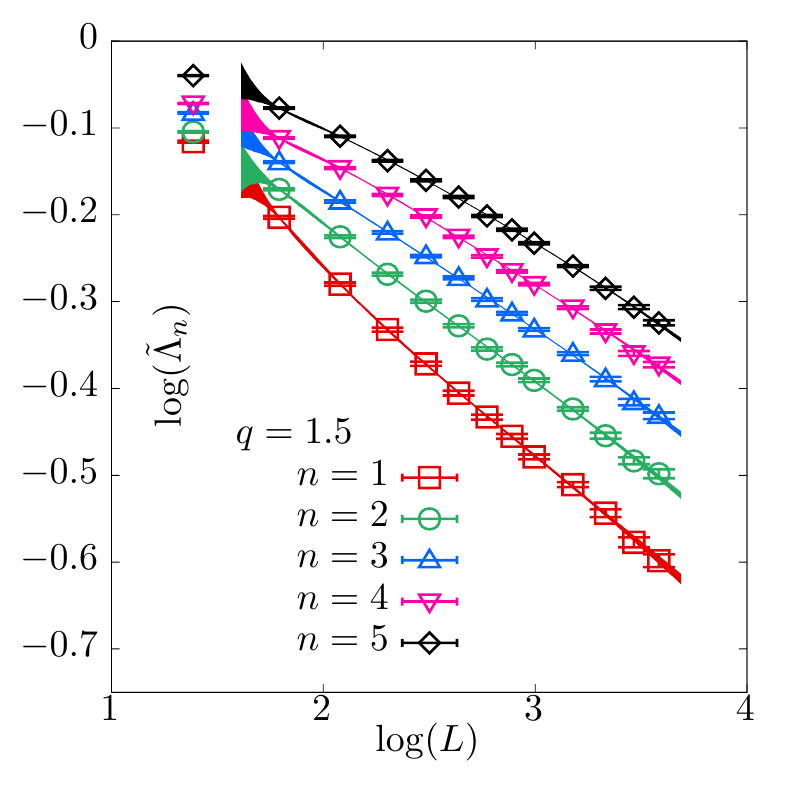}
\includegraphics[scale=0.7]{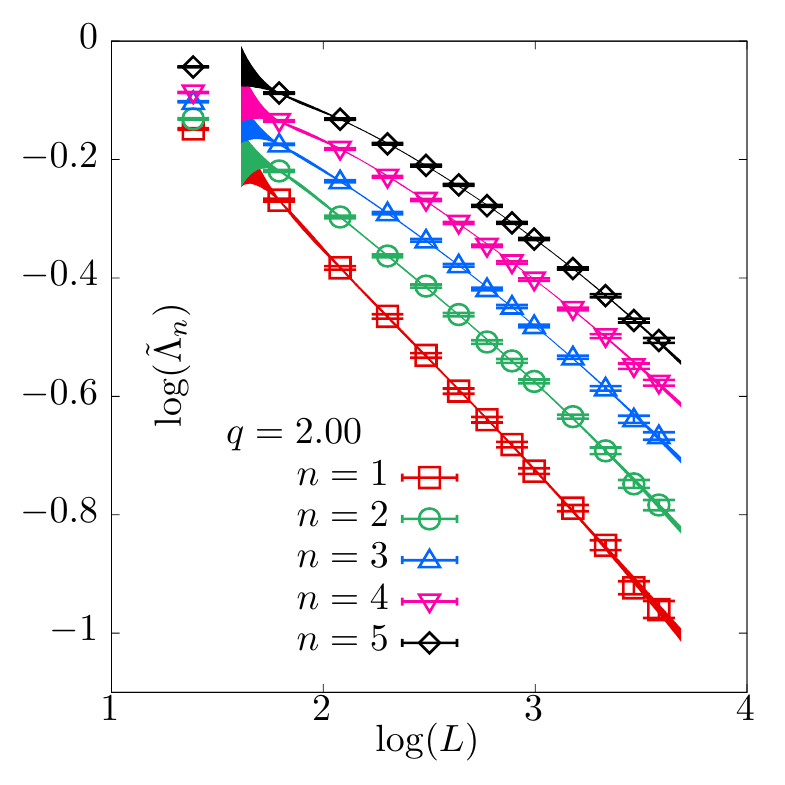}
\includegraphics[scale=0.7]{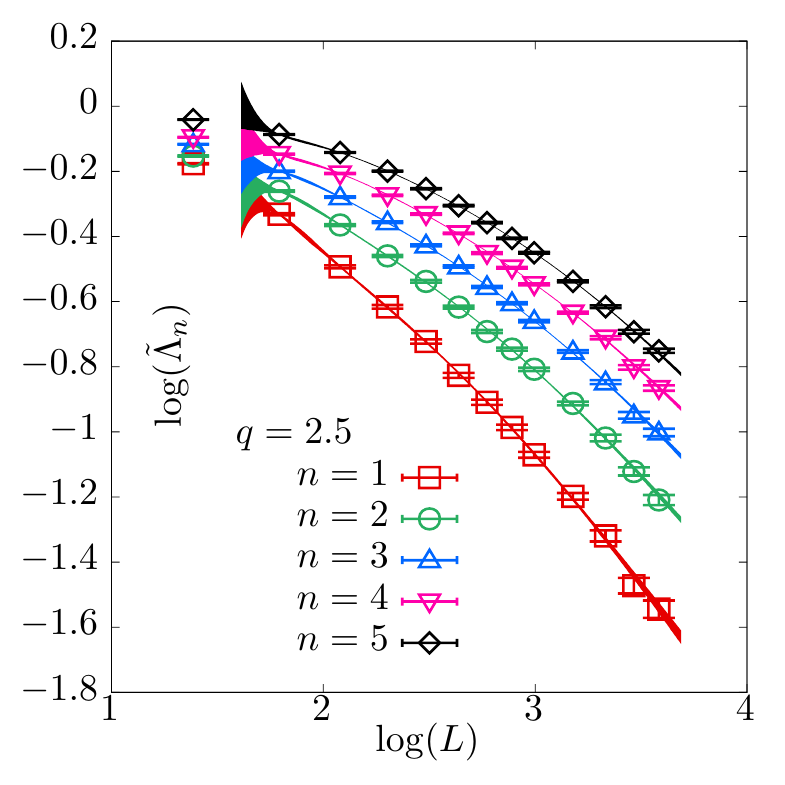}

\includegraphics[scale=0.7]{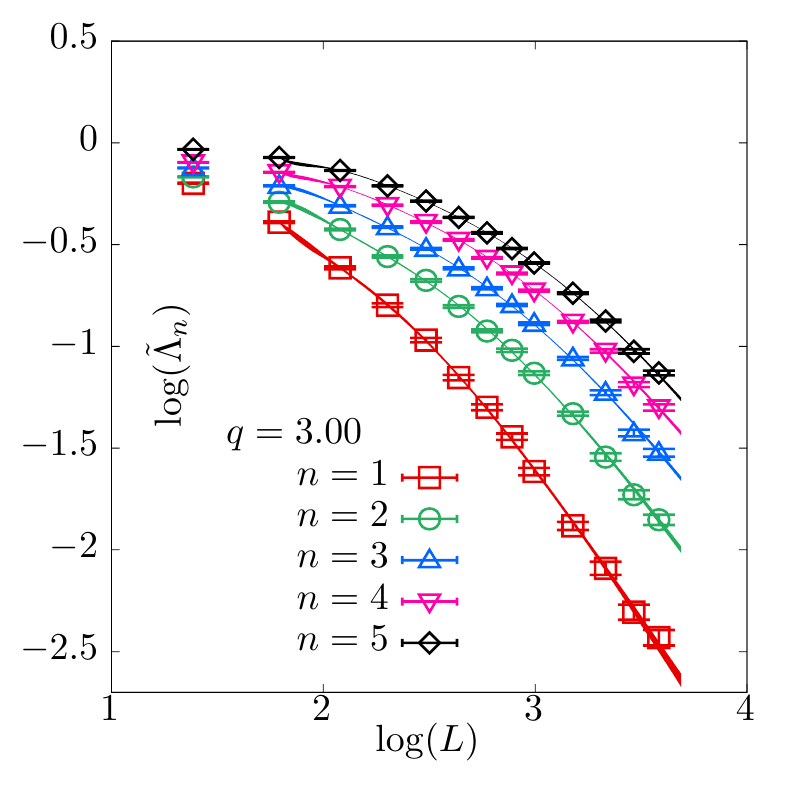}
\caption{
Plot shows $\log(\Lambda_n(L;q)/\Lambda_n(L;q=0)_j)$ versus $\log(L)$
for different charges $q$. In each panel, the different colored
symbols are the first five eigenvalues, $n=1$ to 5.  The eigenvalues are ordered
by magnitude. The curves are the fits as described in the text.
}
\eefs{eigscale}

We determined the low-lying Dirac eigenvalues $\Lambda_i$ with,
$0\le \Lambda_1 \le \Lambda_2\le \ldots$, using the anti-Hermitian
inverse overlap fermion propagator
\beq
{\cal G}^{-2}_q v_i = - \Lambda_i^2 v_i,
\eeq{eigs}
where $v_i$ are the eigenvectors. It is easier determined equivalently using 
\beq
\frac{1}{4}(V(q\theta)+1)(V^\dagger(q\theta)+1) v_i = \frac{\Lambda_i^2}{1+\Lambda_i^2} v_i,
\eeq{eigs2}
using the Kalkreuter-Simma algorithm~\cite{Kalkreuter:1995mm}. We determined the
smallest eight eigenvalues $\Lambda_j$ this way, and used only $j\le 5$ for the
analysis to avoid any inaccuracies in the higher eigenvalues. We
used $q=0.25,0.5,1.0,1.5,2.0,2.5,3.0$ in the eigenvalue studies.
We used $L^3$ lattices with $L=4,6,8,10,12,14,16,18,20,24,28,32,36$.
For each of those $L$ in that order, we used the following number
of configurations; 680,680,680,680,680,680,680,680,680,278,210,153
configurations respectively. We formed the ratio 
\beq
\tilde\Lambda_j=\frac{\Lambda_j(L;q)}{\Lambda_j(L;0)},
\eeq{reducedlambda}
to study the effect of non-zero $q$ and reduce any finite-$L$ corrections already 
present in free theory.

We used the finite-size scaling of the low-lying Dirac eigenvalues
$\tilde\Lambda_i(L)\propto L^{-\gamma_S}$ to determine the scalar
anomalous dimension $\gamma_S$. One way to see it is that the scalar
susceptibility $\chi_q=\int d^3x \langle S^{\pm,0}(x)
S^{\mp,0}(0)\rangle=L^{-3}\sum_j \Lambda_j^{-2}$ scales as
$L^{-1+2\gamma_S}$, which implies that $\Lambda_j\propto L^{-1-\gamma_S}$
for all $j$ in the large-$L$ limit. In \fgn{eigscale}, we have shown
the dependence of $\tilde\Lambda_i$ for $i=1$ to 5 as a function
of $L$ in a log-log scale; the different panels correspond to $q$
ranging from 0.25 to 3.0. One can see that for larger $q$, one does
not a see a perfect $\log(L)$ scaling dependence and the subleading
corrections get larger in the range of $L$ used.  Therefore, we
used the following ansatz to capture the leading $L^{-\gamma_S}$
scaling along with sub-leading corrections which we model to be
$1/L^k$ corrections for integer $k$:
\beq
\tilde\Lambda_j(L)=a_j L^{-\gamma_S}\left(1+\sum_{k=1}^{N_{\rm max}} b_{j,k}L^{-k}\right).
\eeq{scaleansatz}
We performed a combined fit of the above ansatz to the $L$-dependence
of $\tilde\Lambda_j$ for $j=1$ to 5.  Using $N_{\rm max}=4$, we
were able to fit the data at all $q$ ranging from $L=6$ to 36 with
$\chi^2/{\rm dof}<1.2$. The error-bands from such fits are shown
along with the data in \fgn{eigscale}. By reducing $N_{\rm max}=2$,
we were able to fit data ranging from $L=14$ to 36, and there is
possibly a systematic effect to slightly increase the estimated
$\gamma_S$, but such changes were within error-bars.  Therefore,
we take our estimates that fit the data over wider range using
$N_{\rm max}=4$ as our best estimate in this paper. The determinations
of $\gamma_S$ from different ranges of $L$ and goodness-of-fit are
summarized in \tbn{vals}.

\bet
\centering
\begin{tabular}{|c|c|c|c|c|}
\hline
\hline
$q$ & $L$ range & $N_{\rm max}$ & $\gamma_S$ & $\chi^2/{\rm dof}$ \cr
\hline
0.25 & 6-36 &   4 & 0.011(11) & 26.0/34 \cr
     &14-36 &   2 & 0.022(10) & 13.1/24 \cr
     \cline{2-5}
0.50 & 6-36 &   4 & 0.036(21) & 26.5/34 \cr
     &14-36 &   2 & 0.058(15) & 13.9/24 \cr
     \cline{2-5}
1.00 & 6-36 &   4 & 0.112(40) & 28.9/34 \cr
     &14-36 &   2 & 0.156(28) & 17.7/24 \cr
     \cline{2-5}
1.50 & 6-36 &   4 & 0.242(54) & 28.0/34 \cr
     &14-36 &   2 & 0.299(41) & 17.2/24 \cr
     \cline{2-5}
2.00 & 6-36 &   4 & 0.459(68) & 27.8/34 \cr
     &14-36 &   2 & 0.522(56) & 17.6/24 \cr
     \cline{2-5}
2.50 & 6-36 &   4 & 0.888(64) & 32.8/34 \cr
     &14-36 &   2 & 0.922(63) & 22.9/24 \cr
     \cline{2-5}
3.00 & 6-36 &   4 & 1.657(56) & 39.1/34 \cr
     &14-36 &   2 & 1.619(66) & 27.2/24 \cr
\hline
\hline

\hline
\end{tabular}
\caption{The estimates of scalar dimensions $\gamma_S$ as estimated
using the FSS of Dirac eigenvalues $\Lambda_j$ are tabulated. Here,
$q$ is the charge used in the Dirac operator, $L$ range is the range
of lattice size from which the eigenvalues are used in the fits,
$N_{\rm max}$ is the number of $1/L^k$ corrections used in
\eqn{scaleansatz}, and $\gamma_S$ is the estimated scaling dimension
from the fits to \eqn{scaleansatz}. The minimum $\chi^2$ per degree
of freedom of the fits as a measure of goodness-of-fit are also
tabulated.}
\eet{vals}

\section{Two-point functions}
\bef
\centering
\includegraphics[scale=0.8]{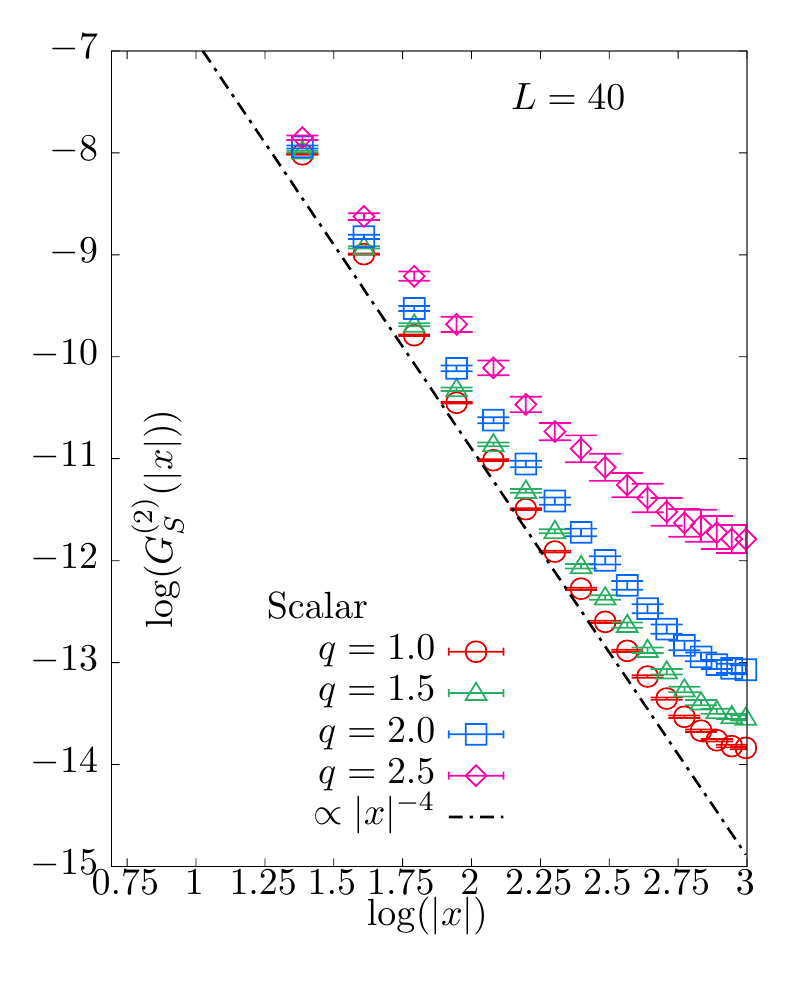}
\includegraphics[scale=0.8]{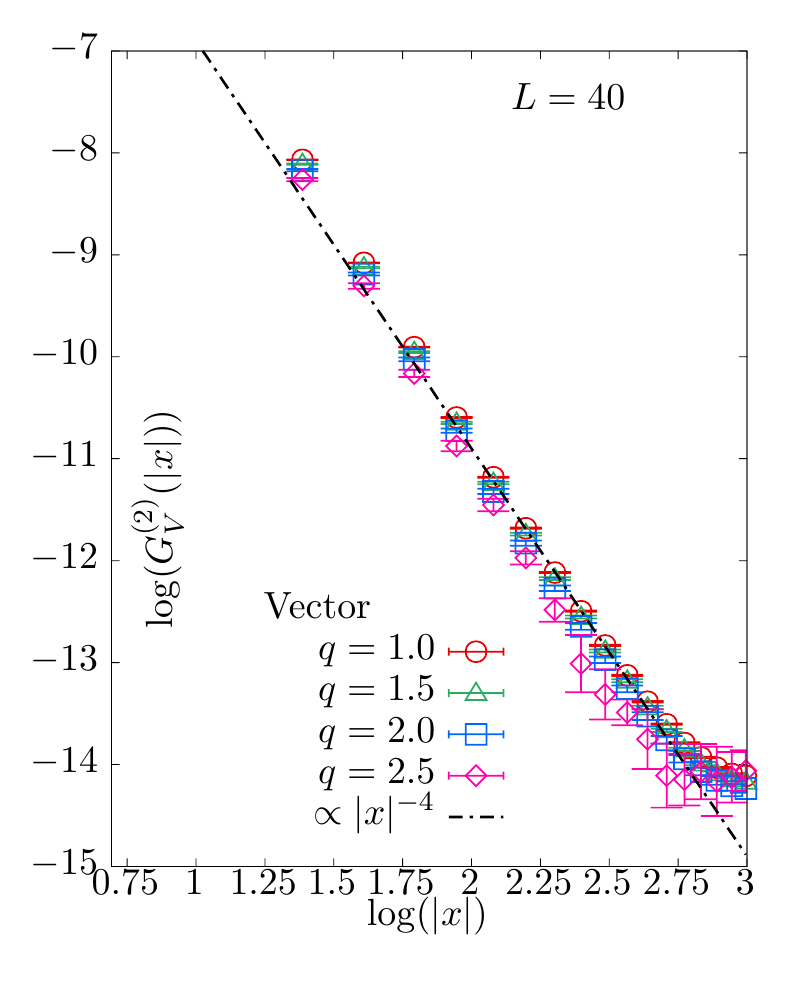}
\caption{
    The flavor triplet scalar and vector correlators from different $q$
    are compared. $L=40$ lattice was used. The conserved vector correlator
    is unchanged up to slight change in normalization $C_V$ that
    seems to decrease with $q$. For the scalar, both the amplitude as well as 
    the exponent $\gamma_S$ changes; the decrease in $\gamma_S$ is also apparent from the plot.
    }
\eef{correls}

\bef
\centering
\includegraphics[scale=0.8]{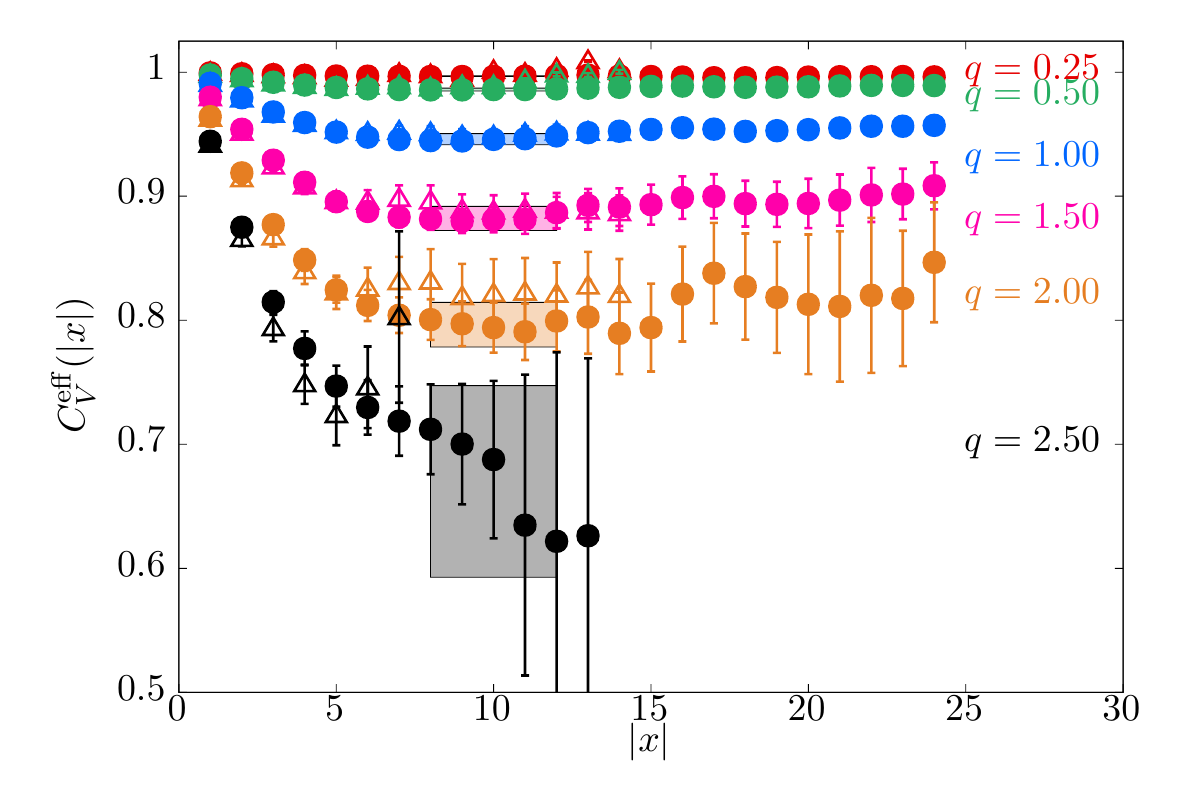}
\caption{
    The effective amplitude $\tilde{C}^{\rm eff}_V(|x|)$, defined in 
    \eqn{effcv}, is shown as a function of operator separation $|x|$. The different 
    colored symbols are from different $q$ specified in right-side of the plot. For each $q$,
    data from $L=28$ (triangle) and $L=48$ (circle) are shown. 
    The value of the vector two-point function amplitude $\tilde{C}_V$ is estimated from
    the plateau region $|x|\in [8,12]$ shown using the bands.
}
\eef{cj}

We computed the two-point functions by coupling fermion sources of
charge $q$ to the gauge fields as described in \eqn{bilin} in the
main text. The expressions for two-point functions in terms of the
fermion propagators are
\beq
G^{(2)}_{S^{\pm,0} S^{\mp,0}}(|x-y|)=\left\langle \sum_{\alpha=1}^2 |{\cal G}^{\alpha\alpha}_q(x,y)|^2\right\rangle;\quad
G^{(2)}_{V^{\pm,0}_\mu V^{\mp,0}_\mu}(|x-y|)=-\left\langle \sum_{\alpha,\beta,\gamma,\delta=1}^2 \sigma_\mu^{\alpha\beta} {\cal G}^{\beta\gamma}(x,y)\sigma_\mu^{\gamma\delta} {\cal G}^{\delta\alpha}(y,x)\right\rangle.
\eeq{2ptexp}
Since all the propagators ${\cal G}$ are determined for the same
value of charge $q$, we have suppressed the index for $q$.  We
determined the correlator in a standard fashion by using a point
source vector $v^{\alpha'}(x')=\delta_{x',x}\delta_{\alpha',\alpha}$
using terms such as
\beq
{\cal G}^{\beta,\alpha}_q(y,x)=\bigg{[}(V-V^\dagger)\cdot\left[(1+V)(1+V^\dagger)\right]^{-1}\cdot v \bigg{]}^\beta(y),
\eeq{cg}
and the identity ${\cal G}^{\alpha,\beta}(x,y)=-{\cal
G}^{\beta,\alpha}(y,x)$ to compute backward propagators. We used
conjugate gradient (CG) to determine
$\left[(1+V)(1+V^\dagger)\right]^{-1}\cdot v$, using a stopping
criterion $3\cdot 10^{-7}$. For the inner-CG to determine $V=(X
X^\dagger )^{-1/2}X$, we used a stopping criterion of $3\cdot
10^{-9}$. We chose $x=(0,0,0)$ and $y=(0,0,z)$ along the lattice
axes, and hence $|y-x|=z$.

In \fgn{correls}, we show the vector and scalar two-point functions
as a function of $|x|$ as determined on $L=40$ lattice. For each
of them, we have shown the correlators from $q=1.0,1.5,2.0$ and 2.5
as the different colored symbols.  For the case of vector which is
a conserved current, the scaling dimension cannot be corrected by
an anomalous dimension and the only change can be its amplitude
$C_V$.  From the plot, one can see a decreasing tendency in $C_V$,
which will analyze in detail in the later part.  For the scalar,
there is both a decrease in scaling dimension due to the non-zero
$\gamma_S$ and a decrease in amplitude.

The correlators on a periodic lattice have three parts; 1) a small
distance part consisting $|x|$ of the order of few lattice spacing
where the operators have contributions from the primary scaling
operators as well as of secondary scaling operators of higher scaling
dimensions. 2) an intermediate $|x|$ that is larger than few lattice
spacings and also smaller than $L/2$ where operator scales with the
scaling dimension of the primary.  3) larger $|x|$ of the order of
$L/2$ where finite size effects take over. In \fgn{correls} for
vector two-point function, we have also shown an expected $|x|^{-4}$
dependence with an appropriately chosen amplitude $C_V$. One can
see that there is only a short intermediate region in $|x|$ on
the typical lattices $L\sim 40$ used, where there is a $|x|^{-4}$
behavior, and hence fitting such a functional form to the correlator
to extract the scaling dimension and the amplitude is not a good
way. Instead, in order to obtain the scaling dimension from the
correlator, it is best to use the finite size scaling; for a critical
theory, the two point function $G^{(2)}(|x|,L)$ should have a scaling
form $L^{-2\Delta}(g(|x|/L)+{\cal O}(1/L))$ and hence by keeping
$|x|=\rho L$ for fixed $\rho$, one can extract $\Delta$ from the
FSS (for example,refer~\cite{Banerjee:2017fcx}). We chose $\rho=1/4$
in the main text.  In order to determine the amplitude $C_V$, we
found it optimal to use the reduced two point function
\beq
\tilde{G}^{(2)}(|x|,q;L)=\frac{G^{(2)}(|x|,q;L)}{G^{(2)}(|x|,q=0;L)},
\eeq{redc2pt}
which removes finite lattice spacing and finite-volume effects that
are already present in free theory; for the vector, which is where
we are interested in the amplitude the most, this was optimal since
the behavior of correlators for non-zero $q$ and zero $q$ were more
of less the same and hence we can find $\tilde{C}_V(q)\equiv
C_V(q)/C_V(q=0)$ very well. For this, we define an effective
$|x|$-dependent $\tilde{C}_V(|x|)$ as
\beq
\tilde{C}^{\rm eff}_V(|x|,q;L)\equiv \sum_{\mu=1}^3 \tilde{G}^{(2)}_{V^+_\mu V^-_\mu}(|x|,q;L).
\eeq{effcv}
If there were perfect $|x|^{-4}$ scaling in both $q\ne 0$ and $q=0$
vector two-point functions, the effective $C_V^{\rm eff}$ will
exhibit a plateau at all distances $|x|$. Instead in the actual
case, one can expect a plateau only over an intermediate $|x|$. In
\fgn{cj}, we show $\tilde{C}_V(|x|,q;L)$ as a function of $|x|$;
the different colors correspond to different $q$ and for each $q$,
we have shown the results using $L=28$ and 48 lattices as the open triangles
and filled circles respectively.  One finds that at fixed $|x|$,
the values of $\tilde{C}_V(|x|,q;L)$ for these ranges of $L$ above 20 are
consistent within errors and hence have reached their thermodynamic
limits within statistical errors.  For $|x|\in [8,12]$ which is
larger than few lattice spacings and at the same time much smaller
than $L/2$ for the values of $L$ used, one finds a plateau and we
estimate $C_V$ by averaging over these values of $|x|$. Such estimates
are shown as the bands in \fgn{cj}. We take the determination of
$C_V$ on the largest $L=48$ we computed to be our estimate. In order
to compute $C_V(q)$, we use the continuum value of
$C_V(q=0)=1/(4\pi^2)$~\cite{Giombi:2016fct}.

\section{Three-point functions}
In a CFT, the conformal invariance dictates the form of three-point
functions of primary operators.  In the lattice model, the local
operators we construct in general are not the scaling operators,
and hence, we expect to observe scaling only when the distances
$|x_{ij}|$ between any pair of operators are large, but at the same
time, smaller than $L/2$.  Therefore, we studied three-point functions
$G_{V^+_iV^-_i V^0_3}, G_{V^+_iV^-_i V^0_3}$ and $G_{S^+_i S^-_i
V^0_3}$ in the main text as further evidence to the conformal nature
of the lattice theory and also to demonstrate that the system is a
very good model system for furthering the lattice framework to study
fermionic CFTs.

We constructed the three-point functions in terms of the fermion propagators
as
\beq
G^{(3)}_{S^+ S^- V^0_\rho}(x_{12},x_{23},x_{31})=-\sqrt{2}{\rm Re}\bigg{[}\left\langle \sum_{\alpha,\beta,\gamma,\delta=1}^2 {\cal G}^{\alpha,\beta}(x_1,x_2) {\cal G}^{\beta,\gamma}(x_2,x_3)\sigma_\rho^{\gamma,\delta} {\cal G}^{\delta,\alpha}(x_3,x_1)\right\rangle\bigg{]},
\eeq{tptssv}
and 
\beq
G^{(3)}_{V^+_\mu V^-_\nu V^0_\rho}(x_{12},x_{23},x_{31})=-\sqrt{2}{\rm Re}\bigg{[}\left\langle \sum_{\alpha,\beta,\gamma,\delta=1}^2 \sigma^{\alpha\beta}_\mu {\cal G}^{\beta\gamma}(x_1,x_2)\sigma_\nu^{\gamma\delta} {\cal G}^{\delta\rho}(x_2,x_3)\sigma_\rho^{\rho\lambda} {\cal G}^{\lambda\alpha}(x_3,x_1)\right\rangle\bigg{]}.
\eeq{tptvvv}
The contractions above require two overlap inversions per choice
of $x_3$ at fixed $x_1$.  From the three-point functions, we
constructed the reduced three-point function
\beq
\tilde{G}^{(3)}_{O^+_i O^-_j O^0_k}\equiv \frac{G^{(3)}_{O^+_i O^-_j O^0_k}(x_{12},x_{23},x_{31};q,L)}{G^{(3)}_{O^+_i O^-_j O^0_k}(x_{12},x_{23},x_{31};q=0,L)}.
\eeq{redtpt}
By construction, in free field theory, this ratio is normalized to
1 at all separations and hence we expect this ratio to remove both
short distance lattice corrections as well as large distance finite
volume corrections.  We specialized to collinear three-point
function~\cite{Osborn:1993cr} in order to greatly simplify the
$x_{ij}$ dependence. That is, we used $x_1=(0,0,0), x_2=(0,0,z_2),
x_3=(0,0,z_2+z_3)$. In this configuration of operators, We expect the
above ratio to behave as
\beqa
\tilde{G}^{(3)}_{O^+_i O^-_j O^0_k}(z_2, z_3)=\tilde{C}_{O^+_i O^-_j O^0_k}z_2^{\gamma_{O_i}+\gamma_{O_j}-\gamma_{O_k}} z_3^{\gamma_{O_j}+\gamma_{O_k}-\gamma_{O_i}} (z_2+z_3)^{\gamma_{O_k}+\gamma_{O_i}-\gamma_{O_j}},
\eeqa{exp1}
where the scaling dimension of the operator is $\Delta_{O}=2-\gamma_O$,
and $\tilde{C}_{O^+_i O^-_j O^0_k}$ is the ratio of OPE coefficients
$\tilde{C}_{O^+_i O^-_j O^0_k}(q)= C_{O^+_i O^-_j O^0_k}(q)/C_{O^+_i
O^-_j O^0_k}(q=0)$. For the vector $\gamma_V=0$. Therefore, the
expressions above simplify for our three-point functions. In general,
if the operators only had overlap with the primary scaling operator,
one would expect,
\beqa
\tilde{G}^{(3)}_{V^+_i V^-_i V^0_3}(z_2, z_3) &=& \tilde{C}_{V^+_i V^-_i V^0_3},\cr
\tilde{G}^{(3)}_{V^+_3 V^-_3 V^0_3}(z_2, z_3) &=& \tilde{C}_{V^+_3 V^-_3 V^0_3},\cr
\tilde{G}^{(3)}_{S^+ S^- V^0_3}(z_2, z_3) &=& \tilde{C}_{S^+ S^- V^0_3} z_2^{2\gamma_S},
\eeqa{exp2}
at all distances. However, on the lattice, our choice of operators
generically overlap with both primary as well as secondary operators,
and hence, one can expect to see the above behavior of three-point
functions with smallest scaling dimensions as that of primaries,
only when
\beq
0 \ll z_2,z_3,z_2+z_3  \ll L/2. 
\eeq{cond1}
Therefore, we turn the expression above around, and define effective 
OPE coefficients $C^{\rm eff}_{O^+_i O^-_j O^0_k}(z_2, z_3)$ as 
\beqa
\tilde{C}^{\rm eff}_{V^+_i V^-_i V^0_3}(z_2,z_3) &\equiv& \tilde{G}^{(3)}_{V^+_i V^-_i V^0_3}(z_2, z_3)  ,\cr
\tilde{C}^{\rm eff}_{V^+_3 V^-_3 V^0_3}(z_2,z_3) &\equiv& \tilde{G}^{(3)}_{V^+_3 V^-_3 V^0_3}(z_2, z_3),\cr
\tilde{C}^{\rm eff}_{S^+ S^- V^0_3}(z_2,z_3)&\equiv& \tilde{G}^{(3)}_{S^+ S^- V^0_3}(z_2, z_3)  z_2^{-2\gamma_S}.
\eeqa{ceffdef}
We presented the results in these effective OPE coefficient in
\fgn{wilcoeff} in the main text.  First, we observe a plateau for
intermediate distances which is a nice demonstration of the spectator
fermion observables satisfying the CFT conditions. We can extract
the OPE coefficient from the value of $C^{\rm eff}$ in the plateau
region.  The condition in \eqn{cond1}, also tells us the optimal
ordering of three-point functions to look at. For example, we could
have constructed a three-point function $\tilde{G}^{(3)}_{S^+ V^-_3
S^0}(z_2, z_3)$ which behaves as $\tilde{C}_{S^+ V^- S^0}
(z_2+z_3)^{2\gamma_S}$. Given a finite $L=64$ lattice we use, if
we used $z_2,z_3\approx 10 \sim  0.16 L$ so as to be in a scaling
region, then $z_2+z_3\approx 20 \sim 0.32 L$, which might suffer
from finite size effect. Hence the usage of infinite-volume factor
$(z_2+z_3)^{2\gamma_S}$ might not be correct. This is the reason,
we used the ordering of operators given in \eqn{exp1} and \eqn{exp2}.

Since the three-point function computation is for a small-scale
computational project we undertook, we used only a single, large
lattice extent $L=64$ and one value of $q=0.15$ using 850 statistically
independent configurations.  We presented the results from this
computation in \fgn{wilcoeff}.

\section{Eigenvalue distributions}
\bef
\centering
\includegraphics[scale=0.7]{his_q10.pdf}

\includegraphics[scale=0.7]{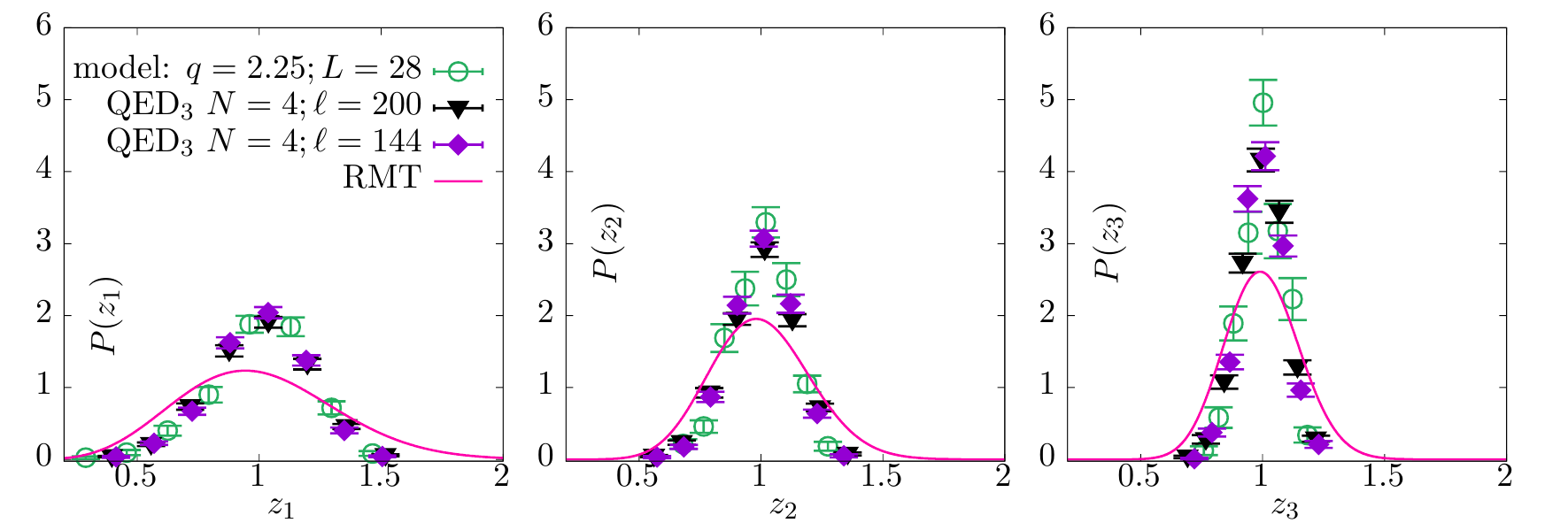}

\includegraphics[scale=0.7]{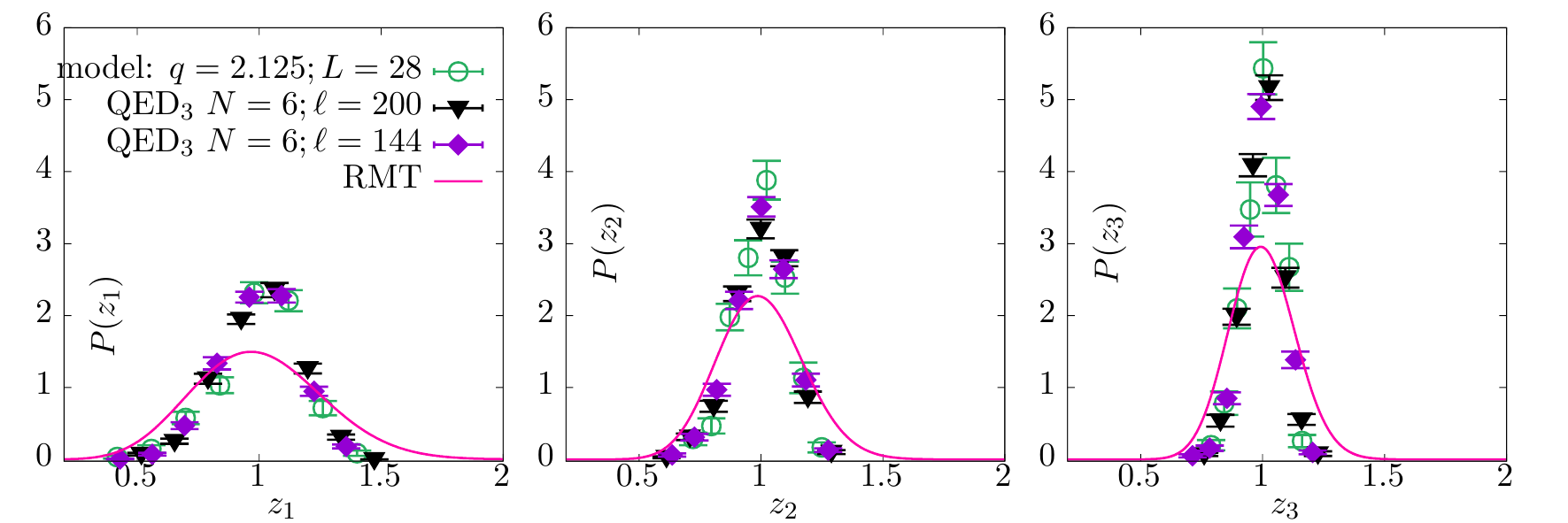}

\includegraphics[scale=0.7]{his_q8.pdf}

\caption{
    Distribution of scaled eigenvalues
    $z_i=\frac{\Lambda_i}{\langle\Lambda_i\rangle}$ for the three
    lowest eigenvalues (left to right) from the conformal lattice
    model at $q=2.5, 2,25, 2.125, 2.0$ (top to bottom) are compared
    with those from $N=2,4,6,8$ QED$_3$ respectively. For the lattice
    model, results from $L=24,28,32$ are shown for $N=2,8$, and
    $L=28$ for $N=4,6$.  For QED$_3$, results from two large box
    sizes $\ell$ (measured in units of coupling $g^2$) are shown.
}
\eef{histoall}

In this section, we compare distributions $P(z_i)$ of scaled Dirac eigenvalues
\beq
z_i\equiv \frac{\Lambda_i}{\langle \Lambda_i \rangle},
\eeq{scleig}
as determined in $N$-flavor QED$_3$ with that in the model using
charge $q$ bilinears.  Here, $q$ is chosen to be in the vicinity
of the expected uncertainty range where we expect $\gamma_S$ for the charge
$q$ bilinear will be equal to that in the $N$ flavor QED$_3$.  As we
mentioned in the main text, we expect these ranges to be
$q\in[2.32,2.76], [1.88,2.49], [1.57,2.03], [1.33,1.88]$ respectively.
We
obtained $P(z_i)$ by histograms of the $z_i$ as sampled in the Monte
Carlo. For QED$_3$, we used different physical boxes of volume
$\ell^3$ (measured in units of Maxwell coupling $g^2$) which we
expect to be in the infrared regime of QED$_3$. For QED$_3$, we
used the eigenvalues data in $L=20,24,28$ (which determines the
lattice spacing, the continuum limit of QED$_3$ is in the limit
$L\to\infty$) from our study using massless Wilson-Dirac
fermions~\cite{Karthik:2015sgq}. Since the results from the three
different $L$ for QED$_3$ gave similar result, we only show the
results for $L=28$ in the histograms below. We also checked that
using eigenvalues from our later study~\cite{Karthik:2016ppr} of
$N=2$ QED$_3$ using exactly massless overlap fermions gave results
consistent with the histograms for $N=2$ QED$_3$ using Wilson-Dirac
fermions shown here.  In \fgn{histoall}, we have compared $P(z_i)$
from $N=2,4,6,8$ from distributions using $q=2.5, 2.25, 2.125, 2.0$
respectively, in the top to bottom panels of \fgn{histoall}. The
results for the lowest three eigenvalues are shown for each $N,q$.
The agreement is almost perfect and supports the claim that the
observed agreement cannot be a mere coincidence.

The $N$-flavor non-chiral Gaussian Unitary Ensemble random matrix
theories (nonchiral RMT)~\cite{Verbaarschot:1994ip} is given by the
partition function
\beq
Z_{\rm RMT}(N)=\int [dH] \det\left(H\right)^N e^{-\frac{1}{2}{\rm tr}(H^2)},
\eeq{rmtdef}
where $H$ are $M\times M$ Hermitian matrices, in the limit of
$M\to\infty$. The eigenvalues $\lambda_i$ in the RMT are the
eigenvalues of $H$ (ordered according to their absolute values).
We compare analytical results~\cite{Nishigaki:2016nka,Damgaard:1997pw}
for the distributions of $\lambda_i/\langle \lambda_i\rangle$ from
the $N$-flavor RMTs in the different panels of \fgn{histoall}.  For
a theory with a condensate, the Dirac eigenvalue distribution must
agree with the one from the corresponding RMT. One can see that the
Dirac eigenvalue distributions disagree with those from the RMT,
and instead, they agree with the distributions from the conformal
gauge theories for tuned values of $q$ studied in this paper.

\section{Comparison with leading $1/N$ results}
Since any remnant $q$-dependent ambiguity in normalization factors
that convert the operators in the model to operators in QED$_3$
cannot affect the scaling exponents, we expect the coefficient of
the leading $q^2$ dependence of $\gamma_S$ to be the same as the
coefficient of $32/N$ in large-$N$ expansion. Indeed, our determination
of the leading coefficient $0.078(11)$ of $q^2$ is consistent with
the large-$N$ expectation~\cite{Chester:2016ref} of $2/(3\pi^2)\approx
0.068$ within errors.  Taking an example $q$-$N$ relation of the
form $32/N=q^2+b q^4+\ldots$, shows that coefficients of orders
$q^4$ and higher might have contributions from all orders in $32/N$,
and hence we do not expect such universality in coefficients beyond
leading order in $q^2$.

A similar comparison for two-point function amplitudes, for example
$C_V$, cannot be made due to possible $q$-dependent conversion
factors, $Z(q)=1+\# q^2+\ldots$, between operators in the model and
in QED$_3$. In fact, we differ in the leading $q^2$ contribution
to $[C_V(q)/C_V(0)-1]$; we find  its $q^2$ coefficient to be
$-0.0478(7)$ where as the coefficient of $32/N$ in large-$N$
expansion~\cite{Giombi:2016fct} for $C_V$ in QED$_3$ is $\approx
0.0045$.  Any such conversion factors have to be determined by
comparing the correlators in the model at a value of $q^2$ with
that in the full $N$-flavor QED$_3$, which is not a very useful
statement as such.  However, one can study renormalization group
invariant concepts such as the degeneracy of current correlators,
as presented in this paper.

\end{document}